\documentclass[5p]{elsarticle}

\usepackage[utf8]{inputenc}
\usepackage{graphicx}
\usepackage[colorlinks,allcolors=cyan!70!black]{hyperref}
\usepackage{lipsum}
\usepackage[T1]{fontenc}

\newcommand{\bra}[1]{\langle#1\rvert} % Bra
\newcommand{\ket}[1]{\lvert#1\rangle} % Ket
\newcommand{\braopket}[3]{\langle #1 | #2 | #3\rangle} % Matrix Element
\newcommand{\expect}[1]{ \langle #1 \rangle} % Expectation value

\usepackage{mathtools, cuted}
\biboptions{sort&compress}

\usepackage{amssymb}
\usepackage{amsmath}
\journal{Chemical Physics Letters}

\begin{document}

\let\today\relax

\makeatletter
\def\ps@pprintTitle{%
  \let\@oddhead\@empty
  \let\@evenhead\@empty
  \def\@oddfoot{\reset@font\hfil\thepage\hfil}
  \let\@evenfoot\@oddfoot
}
\makeatother

\begin{frontmatter}



\title{A quantum picture of light-suppressed photosynthetic charge transfer}


%\affiliation{organization={RIKEN Cluster for Pioneering Research},%Department and Organization
 %           city={Wako},
 %           postcode={351-0198}, 
 %           state={Saitama},
  %          country={Japan}}
  
\author{Guang Yang\corref{cor}} 
\ead{guang.yang@riken.jp}

\author{Gen Tatara} %% Author name

\cortext[cor]{Corresponding author.}

\address{RIKEN Cluster for Pioneering Research, Wako, Saitama 351-0198, Japan}

%% Abstract
\begin{abstract}
We propose a dynamic mechanism for the reversible regulation of photosynthesis in varying light environments. We employ a three-level quantum model to take into account the correlations between charge donors and charge acceptors immediately before photoexcitation, and show that under continuous illumination, the transfer efficiency of a single charge is inversely proportional to the intensity of light, which can be suppressed so severely that it becomes a limiting factor on linear electron transport. This result is used to derive a set of analytical expressions that characterize the light response curves of photosynthetic parameters, including that of gross photosynthetic rate which saturates in high light and has long been assumed to obey a Michaelis–Menten function. We discuss the implications of thermal fluctuation in the light source, and argue that at a given intensity of light, the quantum yields measured with an incandescent lamp may be higher than those measured with a laser, a manifestation of thermal fluctuation in lamp illumination. Our new picture helps understand the observed plastocyanin-dependent electron transport in photosystem~I and provides a donor-side scheme for the onset of irreversible damage to photosystem~II by visible light.
\end{abstract}

%\includegraphics[width=5.5in]{graphicabstract.eps}

%\item The fast and efficient regulation of photosynthetic efficiencies in plants in changing light is often attributed to feedback caused by proton gradient changes across the thylakoid membrane, yet this picture was recently challenged by several experiments.
%\item A quantum picture of the regulatory mechanism is proposed, in which photosynthetic charge transfer is suppressed in high light as a result of the interplay of light-matter interaction and quantum correlations between successive steps down the linear electron transport chain.
%\item The new picture offers a consistent interpretation of the light response data of various photosynthetic parameters as well as the visible-light part of the action spectrum of photodamage.

%% Keywords

\end{frontmatter}



\section{\label{sec:intro}Introduction}

A central topic in plant physiology has been to understand the dynamic regulation and acclimation of photosynthetic machinery in changing environments. Plants in natural habitats, for instance, are subject to fluctuations of ambient light from cloud and canopy covering. To deal with fluctuating light, a fast and efficient control of the yields of photosynthetic reactions is necessary to ensure a regulated electron flow through the functional units and to channel away the excess excitation energy delivered from light-harvesting antenna complexes, thus avoiding congestion and overheating of the whole machinery. Past experiments have indeed demonstrated that photosystem~I (PSI) and photosystem~II (PSII), two functional units where the primary photochemistry takes place, are capable of exceptionally flexible adjustments of photosynthetic efficiencies in response to varying ambient light~\cite{fluctuating-light}. At increasing light intensities, PSI and PSII tune down the chloroplast electron transport in a remarkably concerted fashion and dissipate most of the excess energy as heat. 

\begin{figure}
\centering
\includegraphics[width=3in]{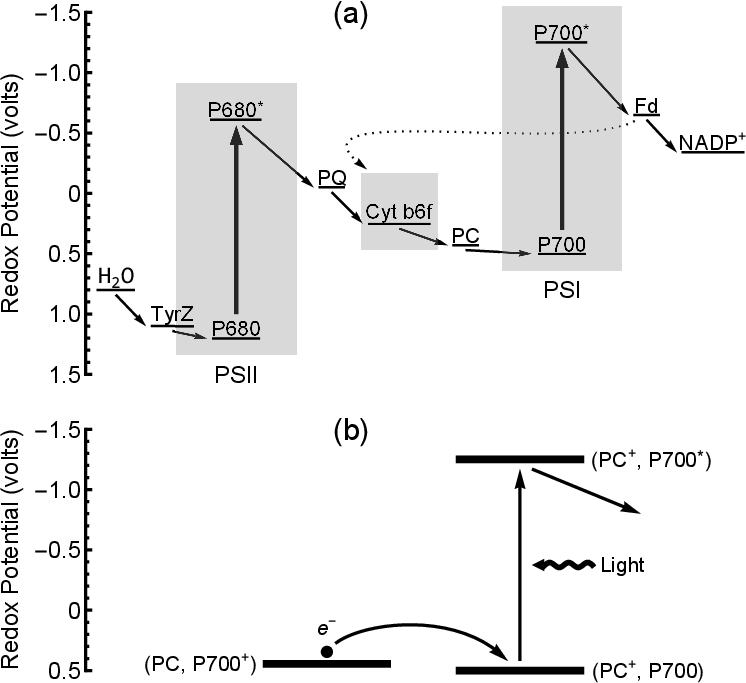}
\caption{(a) Pathways of linear electron flow (solid arrows) through photosystems I and II, via cytochrome $b6f$ complex and assisted by two types of charge carriers plastoquinone and plastocyanin, and cyclic electron flow (dotted arrow) around photosystem I, from ferredoxin back to cytochrome $b6f$ complex. (b) A three-level system as a minimal model to study the effect of photo-blockade, taking photosystem I as an illustrative example, in which an electron may occupy a plastocyanin state, the ground state of P700, or the excited state P$700^*$, corresponding to configuration (PC, P$700^+$), (PC$^+$, P700), or (PC$^+$, P$700^*$), respectively, and will finally relax to outside the system, \emph{i.e.}, to next step in the electron transport chain. Electron transfer (trajectory denoted by thin arrows) is increasingly suppressed at higher intensities of continuous light illumination (thick wavy arrow).}
\label{fig:zero}
\end{figure}

Traditionally, light-induced down-regulation of photosynthesis has been attributed to feedback processes caused by an elevated proton gradient ($\Delta$pH) across the thylakoid membrane~\cite{restriction}. Under higher light conditions, the metabolic consumption of ATP in carbon fixation declines and proton efflux through the ATP synthase slows down, resulting in an increasingly acidic thylakoid lumen. Acidification of the lumen then triggers non-photochemical quenching (NPQ)~\cite{npq} of excessively absorbed light energy and imposes a constriction on the linear electron flow (LEF) at the cytochrome (Cyt) $b6f$ complex located between PSI and PSII, a commonly recognized safety valve in photosynthetic electron transport chain. Accordingly, the yield of PSI and subsequently that of PSII are diminished. In addition to LEF, proton pumping associated with various cyclic electron flow (CEF) pathways around PSI is believed to play a supplemental role in generating $\Delta$pH~\cite{ph-pgr5,ph-pgl1,ph-ndh}. Fig.~\ref{fig:zero}(a) shows the pathways of LEF and CEF, along with the redox potentials at the intermediate steps.

The $\Delta$pH-dependent mechanism of photosynthetic control, despite the general acceptance, is not without questions. Experiments showed that NPQ  saturated at a notably higher light intensity than LEF, indicating the insufficient role of LEF in driving regulatory feedback in high light and the existence of other variables. This was referred to as evidence of a growing portion of CEF in PSI turnover contributing additional $\Delta$pH~\cite{let-question}. A model of CEF-driven regulation of photosynthesis, nevertheless, is somewhat paradoxical, as it suggests simultaneous speed-up and slow-down of electron transport at Cyt~$b6f$~\cite{cet-paradox}. Even more importantly, the $\Delta$pH-dependent mechanism does not explain the observation that the yields of PSI and PSII are both much more sensitive to light environments than NPQ, the modulation of which occurs not only in a fast, dynamic and interconnected fashion,~\cite{shimakawa2018} but starts in low light, well preceding the activation of NPQ~\cite{zivcak2015}. Indeed, a series of recent experiments confirmed that the elevation of $\Delta$pH was negligible in low light,~\cite{huang2018} and cast doubt on the causal relationship between NPQ and electron transport in PSI~\cite{cet-paradox,shimakawa2018,zivcak2015}. This calls for a consideration of other possible scenarios of photosynthetic control in changing light conditions. 

In this paper, we propose a quantum mechanical picture of down-regulation of LEF through photosystems, dubbed photo-blockade. We employ a simple model of three quantum levels to account for the essential steps in electron transport before and during photoexcitation, and show that the interplay of quantum correlations between these levels and light–matter interaction can lead to a decline in the forward transfer rate of an electron, the extent of which is proportional to the intensity of light. A motivation of our modeling is that excitonic energy transfer from the antenna complex to the reaction center is a fast and quantum mechanically coherent process, which allows to represent the collectively excited chromophores in the antenna and in the reaction center by a single quantum level to a first approximation. This is viable when we restrict our analysis to the level of phenomenology, considering also that the energy transfer process is separable from the much slower charge transfer processes into and within the reaction center, as suggested in experiments~\cite{haehnel1984,chenu2015,debate1}. Our approach may be thought of as drawing upon the  ``trap-limited'' model of exciton kinetics,~\cite{holzwarth2003} in which the rate of primary charge separation is slower than that of energy transfer from the antenna to the reaction center especially in natural conditions. We refrain, however, from commenting on previous transient absorption and fluorescence measurements, in particular because here we study photosystems under steady illumination, a condition different from that in the experiments.
We will discuss in Section~\ref{sec:conclusion} how our model may be extended to gain insights at a more microscopic level. 
Focusing on the elementary process of photoexcitation, we adopt a quantized description of light and assume that different photon modes contained in a continuous spectrum of light (\emph{e.g.,} sunlight) contribute independently to the reaction. We also assume that the system couples strongly to the action of light, motivated by a recent spectroscopic study suggesting that charge excitations in the reaction center possess unusually large dipoles~\cite{novoderezhkin2007}. Our model may be considered a minimal construction for discussing sequential reactions involving photon absorptions. We will show that in nanoscale systems where quantum effects are non-negligible and light–matter interactions are strong, suppression of reactions prior to a photoexcitation is a general trend in continuous high light, regardless of the properties of the light, \emph{e.g.,} the degree of coherence.
This finding may prove useful in understanding charge transfer data in realistic photosystems. For instance, it implies that electron tunnelings from plastocyanin (PC) and tyrosine Z (TyrZ) to reaction center chlorophylls P700 in PSI and P680 in PSII, respectively, are both suppressed at increasing irradiance. This effect, photo-blockade, offers an explanation of the coordinated responses of PSI and PSII to light changes, and possibly serves as a mechanism of photoprotection preventing passage of excess excitations. In very low light in the presence of an abundance of ground-state P700 and P680, absorptions of photons are independent and unconstrained events. This may change as light grows stronger where both PSI and PSII undergo a transition from unconstrained to constrained photoexcitation. In the latter case, rates of charge separations at PSI and PSII are limited by the reduction rates of oxidized P700 and P680 (P$700^+$ and P$680^+$), respectively, which in turn are governed by photo-blockade. 
Likely signatures of such a transition in the nature of photoexcitation were recently reported~\cite{miyake2020}. In this picture, down-regulation of photosynthesis is a direct action of light, rather than one depending on feedback processes. 

The picture of photo-blockade sheds light on the observed PC-dependent transport in PSI,~\cite{haehnel1982} and allows for a consistent interpretation of experimentally measured light responses of photosynthetic parameters~\cite{klughammer1994-PSI,klughammer2008-PSI,krause1991,baker2008-PSII}. In particular, it provides theoretical evidence for the empirically established Michaelis–Menten type light response curve of gross photosynthetic rate, and assigns physical significance to the convexity of the curve~\cite{thornley1976}. 
For PSII, it provides a donor-side scheme for the induction of long-lived P$680^+$ by visible light causing irreversible damage,~\cite{jones1966,santabarbara2001} complementary to a recent theory that addressed  the ultraviolet part of the action spectrum of photodamage~\cite{hakala2005,ohnishi2005,allakhverdiev2005}.

The workings of photo-blockade rely on the availability of pathways to dissipate excess excitation energy. 
We suggest that photo-blockade acts as a zeroth-order regulator of photochemistry, whereas the $\Delta$pH-dependent processes play the vital roles of an energy dissipator, an LEF fluctuation sensor and an overall redox homeostasis stabilizer, ensuring balanced charge and energy transfer in chloroplasts. In addition, several other feedback processes are essential  in low light in the absence of a large enough $\Delta$pH to trigger NPQ, including charge recombination,~\cite{ph-pgr5,zivcak2015} CEF around PSI,~\cite{ph-pgr5,ph-pgl1,ph-ndh} light-induced size changes of granal thylakoid lumen,~\cite{kirchhoff2011} and phosphorylation-enhanced energy exchange between PSI and PSII~\cite{grieco2015}.

This paper is organized as follows. In Section~\ref{sec:model}, we model photo-blockade in an isolated system and in an open system interacting with an environment. The results are used in Section~\ref{sec:discussion} to interpret the light response data of photosynthetic parameters. We propose experiments to test our model, and conclude with a discussion of the limitations and possible extensions of the model in Section~\ref{sec:conclusion}.

\section{\label{sec:model}Model}

We consider electron turnover on the donor sides of PSI and PSII, where electrons are transferred from the immediate donors, PC and TyrZ respectively, to P700 and P680, and then are excited upon absorption of photons to create excited states P$700^*$ and P$680^*$ which further pass on the electrons to form initial charge separations. The electronic system of interest at PSI is a three-level one, illustrated in Fig.~\ref{fig:zero}(b), with an electron possibly occupying a PC state, denoted by a quantum state $\ket{L}$ and corresponding to the configuration (PC, P$700^+$), or occupying the ground state of P700, denoted by $\ket{R}$ and corresponding to the configuration (PC$^+$, P700), or occupying the excited state P$700^*$, denoted by $\ket{R^{*}}$ and corresponding to the configuration (PC$^+$, P$700^*$). Here PC$^+$ stands for oxidized PC. In reality P$700^*$ may be understood as a superposition state of excitons, and the three-level system may be imagined a charge transfer complex formed by PC and PSI. The configurations keep track of the association of the electron with the molecules in the complex, following usual notation~\cite{haehnel1996}. Similarly defined configurations are implied for PSII when needed.

The system under consideration is not in thermal equilibrium but is one driven by light. To model the dynamic effects of light, we single out the photon mode  most resonant with the electronic transition and adopt a full quantum mechanical description of its interactions by a Jaynes–Cummings term~\cite{JC-shore}. 
A similar idea is often employed in the lab, where a monochromatic light is used in photosynthetic measurements to rule out potential mixing of undesirable signals. 
The Hamiltonian is 
\begin{equation}
H=\left( \begin{array}{ccc}
\Delta_1 & \Gamma & 0\\
\Gamma & 0&g a^{\dagger}\\
 0 & g a &\Delta_2
 \end{array} \right) + \omega a^{\dagger} a
  \label{eq:one},
 \end{equation}
where $a^{\dagger}$ is the photon creation operator, $\omega$ is the photon energy, and $g$ is the electron–photon coupling strength. $\Delta_1$ and $\Delta_2$ are the energy differences between $\ket{L}$ and $\ket{R}$ and between $\ket{R}$ and $\ket{R^{*}}$, respectively. $\Gamma$ is the experimentally tunable transition amplitude between $\ket{L}$ and $\ket{R}$ in the redox reaction~\cite{marcus1985}. In PSI photochemistry, for instance, application of PC inhibitors effectively reduces the statistical average of $\Gamma$~\cite{haehnel1982}. We set $\hbar=1$ throughout the paper.

In what follows we study this Hamiltonian, first in an isolated system and then in an open system subject to environment-induced relaxation. In an isolated system, the total population of the electron is unity and the local charge is conserved. In an open system, the electron eventually escapes from the system via decay from $\ket{R^*}$ into the environment. 
Later when we discuss steady-state transport, we assume in addition that electrons are fed in $\ket{L}$ one after another by LEF at a rate independent of light intensity, once and only if the previous electron has left the system. This feeding rate should not be confused with the escape rate of a newly fed electron from the open system, which as we will show declines at increasing light intensities. 
In reality, feeding of electrons to PSI is achieved through unbinding of PC$^+$–P$700^+$ pairs after photoexcited electrons are transferred away, and then binding of new free PC to form new charge transfer complexes (\emph{i.e.}, new three-level systems). Importantly, when an PC$^+$ still occupies the binding site, binding of a new PC is not possible, and direct electron donation to P$700^+$ from free PC outside the complex  is negligible~\cite{haehnel1996}. Two electrons cannot be transferred at the same time through the same site. 
Also, the binding/unbinding process occurs orders of magnitude faster than the measurement of photosynthetic light responses~\cite{klughammer1994-PSI}. If steady-state electron turnover is mainly limited by the duration the electron is in the complex, the local charge may still be viewed as a conserved quantity. We thus expect that the analysis of an isolated system provides useful information to be compared with light response data in steady state.

Owing to the existence of a large redox potential gradient up the electron transport chain, Fig.~\ref{fig:zero}(a), we assume negligible back-tunneling of electrons from $\ket{L}$. Thus, the flow of electrons is largely one-way and we focus on studying time evolution of the system with an electron initially in $\ket{L}$. 
Deviations of realistic transport from the description of a single electron model with a conserved local charge will be discussed in Section~\ref{sec:discussion}.

\subsection{\label{sec:isolated}Isolated system: suppressed quantum tunneling}

Consider an isolated system described by the Hamiltonian in Eq.~(\ref{eq:one}). We are interested in the population dynamics of an electron initially occupying $\ket{L}$. Three different initial states of photons are included in the analysis, a Fock state, a coherent state, and a thermal state. A Fock state enjoys the mathematical simplicity of a fixed photon number and serves as a comparative example. A coherent state is an idealized quantum description of a laser. A thermal state, or a ``chaotic state'', contains a considerable amount of thermal fluctuation and is representative of an optically filtered incandescent lamp sometimes used in the experiments.
We assume that the system is initially separable,~\cite{werner1989} \emph{i.e.}, at $t=0$ the system is described by a product of the density matrices of the electron and the photons, $\rho_{\textrm{tot}} (0)= \rho (0)\otimes \rho_{\textrm{ph}}(0) $, where the electron density matrix $\rho (0) = \ket{L}\bra{L}$. At time $t$, the total density matrix $\rho_{\textrm{tot}} (t) = e^{-iH t}\rho_{\textrm{tot}}(0)e^{iH t}$. 

\subsubsection{\label{sec:fock}Fock state}

We first study a Fock state of $n$ photons at $t=0$, described by $\rho_{\textrm{ph}}(0) = \ket{n}\bra{n}$. We define the shorthand notation $\ket{L, n} = \ket{L}\otimes\ket{n}$.
The Hamiltonian can be cast in a block-diagonal form in the Hilbert space spanned by $\ket{L,n}$, $\ket{R,n}$, and $\ket{R^{*},n-1}$,
  \begin{equation}
H_n=\left( \begin{array}{ccc}
\Delta_1 & \Gamma & 0\\
\Gamma & 0&g_n\\
 0 & g_n &\Delta
 \end{array} \right) + n \omega \mathbf{1}_{3\times 3}
   \label{eq:two},
 \end{equation}
where $g_{n} = g \sqrt{n}$, and $\Delta = \Delta_2-\omega$ is the energy mismatch between incident photons and the electronic transition, sometimes referred to as the detuning. The constant term $n \omega \mathbf{1}_{3\times 3}$ contributes a phase in the wave functions but disappears in the density matrices. We neglect this term henceforth, which amounts to working in the interaction picture.

\begin{figure}
\centering
\includegraphics[width=3in]{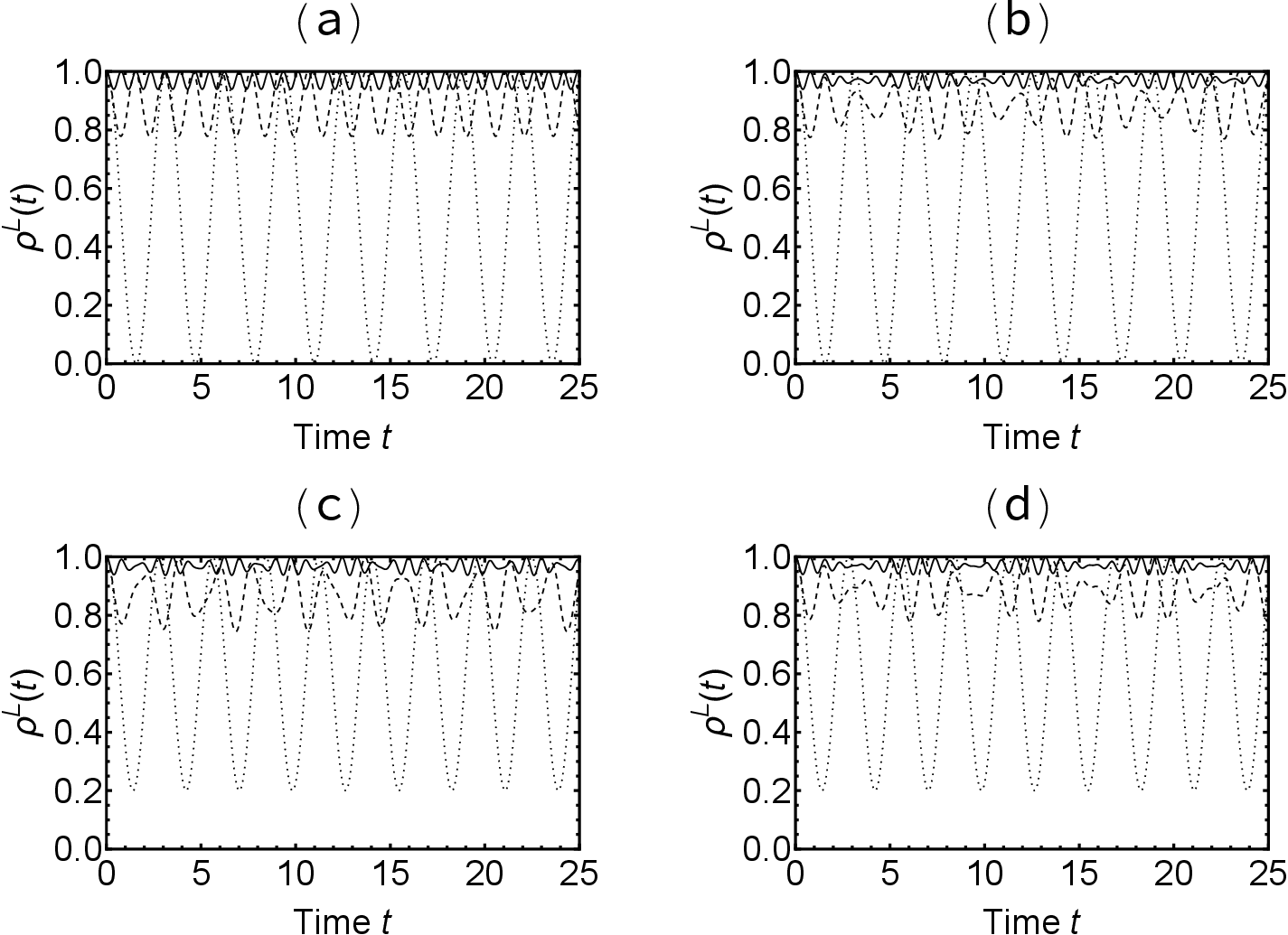}
\caption{Time-dependent population in $\ket{L}$ of an electron interacting with a Fock state of $n=64$ (solid curves), $n=16$ (dashed curves) and $n=0$ (dotted curves) photons in an isolated system. The parameters are $\Delta_1 =0$ and $\Delta =0$ in (a), $\Delta_1 =0$ and $\Delta =1$ in (b), $\Delta_1 =1$ and $\Delta =0$ in (c), and $\Delta_1 =1$ and $\Delta =1$ in (d). In all the plots $g=\Gamma=1$}.
\label{fig:one}
\end{figure}

Let $\ket{\Psi_n(t)} = e^{-iH_n t}\ket{\Psi_n(0)}$ be the total wave function at time $t$, where $\ket{\Psi_n(0)}= \ket{L,n}$. Then $\rho_{\textrm{tot}} (t) =\ket{\Psi_n(t)}\bra{\Psi_n(t)}$. The reduced density matrix of the electron is obtained by tracing out the photon field, $\rho(t)=\text{Tr}_{\textrm{ph}} \rho_{\textrm{tot}}(t) $, whose diagonal elements $\rho^L(t)$, $\rho^R(t)$, $\rho^{R^*}(t)$ give the probabilities of finding the electron in $\ket{L}$, $\ket{R}$, $\ket{R^*}$, respectively, at a given time. In Fig.~\ref{fig:one}, we show numerically calculated time-dependent population in $\ket{L}$ for different parameter settings (see Appendix~A in the Supplementary Materials for the populations in $\ket{R}$ and $\ket{R^*}$). We find that in general, quantum tunneling of the electron from the initial state $\ket{L}$ is suppressed at large photon numbers. This holds true regardless of the choices of the redox potential difference $\Delta_1$ and the detuning $\Delta$. Moreover, the suppression effect is most prominent when the photons are on resonance and diminishes monotonically at increasing detunings (not shown). In the limit of a very large detuning, the excited level $\ket{R^*}$ may be adiabatically eliminated, and suppression of tunneling from $\ket{L}$ to $\ket{R}$, relying on light-induced transitions, becomes less effective.

We are thus motivated to invoke an analytical approach to a better understanding of this localization effect. For simplicity and inspired by the estimated small redox potential difference between PC and P700 within the charge transfer complex,~\cite{haehnel1996} we assume degenerate $\ket{L}$ and $\ket{R}$, and solve the Schr\"{o}dinger equation to obtain a solution of the form (see Appendix~B in the Supplementary Materials for details), 
\begin{equation}
\ket{\Psi_n(t)} =c_{1,n}(t)\ket{L,n}+c_{2,n}(t)\ket{R,n}+c_{3,n}(t)\ket{R^{*} ,n-1}
\label{eq:three},
\end{equation}
with which we find the analytical expressions of density matrices. 
A further simplification of the results is rendered with the assumption of a zero detuning, \emph{i.e.}, the incident photons are resonant with the electronic transition from $\ket{R}$ to $\ket{R^*}$. The eigenvectors of $H_n$ are then found to be
\begin{align}
\ket{0,n}&= \cos{\theta_n} \ket{L,n} -\sin{\theta_n} \ket{R^*,n-1}, \nonumber \\
\ket{+,n}&= \frac{\sin{\theta_n}}{\sqrt{2}}  \ket{L,n} + \frac{1}{\sqrt{2}} \ket{R,n} + \frac{\cos{\theta_n}}{\sqrt{2}} \ket{R^*,n-1}, \nonumber \\
\ket{-,n}&= \frac{\sin{\theta_n}}{\sqrt{2}}  \ket{L,n} - \frac{1}{\sqrt{2}} \ket{R,n} + \frac{\cos{\theta_n}}{\sqrt{2}}  \ket{R^*,n-1}
\label{eq:four},
 \end{align}
 with eigenvalues $0$, $\lambda_n$, $-\lambda_n$, respectively, where $\tan{\theta_n}=\Gamma/g_n$ and $\lambda_n = \sqrt{\Gamma^2+g_n^2}$. The eigenvector $\ket{0,n}$, having a vanishing $\ket{R}$ component due to destructive quantum interference, is a ``dark state'' ~\cite{quantum-optics}. (Note its difference from a symmetry-forbidden state,~\cite{pariser1956-dark} \emph{e.g.}, a carotenoid ``dark state''.) 

We rewrite the initial total wave function in terms of the eigenvectors,
 \begin{equation}
\ket{\Psi_n(0)}= \cos{\theta_n} \ket{0,n} + \frac{\sin{\theta_n}}{\sqrt{2}}\Big(\ket{+,n}+\ket{-,n}\Big)
\label{eq:five},
\end{equation}
to find its time evolution.
At time $t$, the total wave function has the form of Eq.~(\ref{eq:three}), with 
\begin{align}
c_{1,n}(t)&= \cos^2{\theta_n} + \sin^2{\theta_n}\cos{\lambda_n t}, \nonumber \\
c_{2,n}(t)&= - i \sin{\theta_n}\sin{\lambda_n t}, \nonumber \\
c_{3,n}(t)&= \sin{\theta_n}\cos{\theta_n}(\cos{\lambda_n t}-1)
\label{eq:six},
 \end{align}
 and the diagonal matrix elements $\rho^L(t)=|c_{1,n}(t)|^2$, $\rho^R(t)=|c_{2,n}(t)|^2$, and $\rho^{R^*}(t)=|c_{3,n}(t)|^2$. In Eq.~(\ref{eq:six}), the only time dependence is through $\cos{\lambda_n t}$ (or $\sin{\lambda_n t}$). Thus, we may obtain time-averaged populations over a single oscillatory period $ [0,2\pi/\lambda_n ]$, by setting $\overline{\cos{\lambda_n t}}=0$ and $\overline{\cos^2{\lambda_n t}}=1/2$. Let $P_L=\overline{\rho^L(t)}$, $P_R=\overline{\rho^R(t)}$, and $P_{R^*}=\overline{\rho^{{R^*}}(t)}$. We find 
 \begin{align}
P_L&=\cos^4{\theta_n}+ \frac{1}{2} \sin^4{\theta_n}, \nonumber \\
P_R& = \frac{1}{2}\sin^2{\theta_n}, \nonumber \\
P_{R^*}& = \frac{3}{2} \sin^2{\theta_n}\cos^2{\theta_n}
\label{eq:seven}.
 \end{align}
It is easy to verify that $P_L+P_R+P_{R^*}=1$, \emph{i.e.}, the total population is conserved.
 
We are interested in the ``leaked population'' out of $\ket{L}$, defined as $\delta P_L = 1-P_L$, as a measure of the suppression of quantum tunneling. Using $\sin^2{\theta_n}=\Gamma^2/(\Gamma^2 + g_n^2)$, we find $\delta P_L=\delta P_{L;1}-\delta P_{L;2}$, where the first- and second-order contributions
 \begin{align}
\delta P_{L;1} &=\frac{2}{1+\epsilon n}, \nonumber\\
\delta P_{L;2} &= \frac{3/2}{(1+\epsilon n)^2} 
\label{eq:eight},
\end{align}
with $\epsilon = g^2/\Gamma^2$. As a quick check, at $n=0$ we have $\delta P_L = 1/2$, and thus there is $50\%$ probability that the electron is in $\ket{L}$. This is certainly true, because when there is no photon in the system, $\ket {R^*}$ decouples from $\ket {L}$ and $\ket {R}$, and the electron oscillates between $\ket {L}$ and $\ket {R}$ with equal probabilities. On the other hand, as $n \rightarrow \infty $, $\delta P_L \rightarrow 0$, \emph{i.e.}, the electron is permanently locked in $\ket {L}$. This is again expected, because in this case $\sin{\theta_n}\rightarrow0$, and the initial state $\ket{\Psi_n(0)}= \ket{L,n}$ approaches the dark state $\ket{0,n}$ in Eq.~(\ref{eq:four}). Thus, the electron will remain in the initial state indefinitely. We emphasize that the occurrence of suppression of quantum tunneling does not depend on the convenient mathematical conventions chosen here, as shown in Fig.~\ref{fig:one}, or the realization of a true dark state which is very unlikely in natural environments.

The new parameter $\epsilon$ is related to the intensity of incident light. To see this, notice that $g$ is in fact the polarization energy due to coupling of the molecular dipole with the electric field carried by one incident photon,~\cite{quantum-optics}
\begin{equation}
g = - \vec{\xi}\cdot \vec{d} \sqrt{ \frac{ \omega}{2 \mu_0 V} }   
\label{eq:nine},
\end{equation}
where $\vec{\xi}$ is the polarization vector of the photon, $\vec{d} =\braopket{R}{\hat{\vec{d}}}{R^*}$ is the dipolar interaction matrix element assumed to be real without loss of generality, $\mu_0$ is vacuum permittivity, and $V$ is the size of the system. The electric field strength is given by the amplitude of canonically quantized field in the Coulomb gauge,
\begin{equation}
\vec{\mathcal{E}} =\vec{\xi}  \sqrt{ \frac{ \omega}{2 \mu_0 V}} \left( a  + a^{\dagger}  \right)
\label{eq:ten},
\end{equation}
in the long-wavelength limit assumed for a bound-state electron in the Jaynes–Cummings model. At the same time, $\Gamma$ defined in Eq.~(\ref{eq:one}) also has the dimension of energy, $\Gamma= \braopket{L}{\hat{A}}{R}$, where $\hat{A}$ is the tunneling operator between $\ket{L}$ and $\ket{R}$. Thus, 
\begin{equation}
\epsilon =  \frac{ |\braopket{R}{\vec{\xi} \cdot \hat{\vec{d}}}{R^*}|^2}{|\braopket{L}{\hat{A}}{R}|^2} \mathcal{E}^2 
\label{eq:eleven}
\end{equation}
is a dimensionless quantity proportional to the squared one-photon electric field strength and inversely proportional to the system size. In the classical picture, $\mu_0 \mathcal{E}^2$ is the energy density of an electromagnetic wave. Thus, heuristically, $\epsilon$ may be interpreted as proportional to the energy density of the electromagnetic radiation created by a single photon, and $\epsilon n$ is proportional to the total energy density of all incident photons, \emph{i.e.}, the macroscopic intensity of light. In an optical cavity of a finite size, photon number is a meaningful quantity. In free space such as a natural environment, however, $\epsilon n \propto n/V$ is physically more significant. We define the macroscopic limit
\begin{equation}
n  \rightarrow \infty, \;  \;  \epsilon \rightarrow 0, \;   \;  \epsilon n  \rightarrow \text{constant}
 \label{eq:twelve}.
\end{equation}
Eq.~(\ref{eq:eight}) thus describes the suppression of quantum tunneling as a function of the macroscopic light intensity. In high light with $\epsilon n \gg1$, the contribution of $\delta P_{L;2}$ is negligible, and $\delta P_L\rightarrow\delta P_{L;1}$. In fact, using $\sin^2{\theta_n}=1/(1+\epsilon n)$, we see that the steady-state populations in Eq.~(\ref{eq:seven}) are all macroscopic quantities. 

By Eq.~(\ref{eq:eleven}), $\epsilon n$ is determined not only by the absolute light intensity, but relative to the bare parameter $\Gamma$. We will refer to it as the relative light intensity.

\subsubsection{\label{sec:coherent}Coherent state}

Next, we study quantum tunneling of the electron assuming that the photons are initially in a coherent state described by $\rho_{\textrm{ph}}(0) = \ket{\alpha}\bra{\alpha}$, where 
\begin{equation}
\ket{\alpha}  = e^{-\frac{\alpha^2}{2}} \sum_{n=0}^{\infty} \frac{\alpha^n}{\sqrt{n!}} \ket{n} 
 \label{eq:thirteen}
\end{equation}
is the coherent state wave function in the photon-number basis. 
We have assumed without loss of generality that $\alpha$ is real. The probability of finding $n$ photons in $\ket{\alpha}$ follows a Poisson distribution 
 \begin{equation}
p_{\alpha,n}=e^{-m_{\alpha}} \frac{(m_{\alpha})^n}{n!} 
 \label{eq:fourteen},
\end{equation}
where $m_{\alpha} = \alpha^2$ is the mean photon number. 

As in the treatment of a Fock state, we assume $\Delta_1=\Delta=0$ and rewrite the initial total wave function $\ket{\Psi(0)}=\ket{L,\alpha}$ in terms of the eigenvectors in Eq.~(\ref{eq:four}) to find its time evolution. The total wave function at time $t$ is
 \begin{equation}
\ket{\Psi(t)}=e^{-\frac{\alpha^2}{2}} \sum_{n=0}^{\infty}  \frac{\alpha^n}{\sqrt{n!}} \ket{\Psi_n(t)}
 \label{eq:fifteen},
\end{equation}
where $\ket{\Psi_n(t)}$ is defined in Eq.~(\ref{eq:three}) and the coefficients are given in Eq.~(\ref{eq:six}). We then have $\rho(t)=\text{Tr}_{\textrm{ph}} \ket{\Psi(t)}\bra{\Psi(t)} $.
Upon performing time average on $\rho^L(t)$, $\rho^R(t)$, and $\rho^{R^*}(t)$, we find the leaked population out of $\ket{L}$, $\delta P_L=\delta P_{L;1}-\delta P_{L;2}$, with
\begin{align}
\delta P_{L;1} & = \sum_{n=0}^{\infty} p_{\alpha,n}  \frac{2}{1+\epsilon n}, \nonumber \\
\delta P_{L;2} & = \sum_{n=0}^{\infty} p_{\alpha,n}  \frac{3/2}{(1+\epsilon n)^2} 
\label{eq:sixteen}.
\end{align}
These results differ from those in Eq.~(\ref{eq:eight}) for a Fock state by a summation over photon number, weighted by the Poisson distribution $p_{\alpha,n}$ defining the coherent state. In Appendix~C of the Supplementary Materials, we derive the compact expressions of $\delta P_{L;1}$ and $\delta P_{L;2}$ as functions of $\epsilon$ and $m_{\alpha}$. 
Numerical calculation (not shown) confirms that $\delta P_{L;2}$ becomes vanishingly small at large photon numbers $m_{\alpha} \gg 1$ for fixed $\epsilon$, or in high light $\epsilon m_{\alpha} \gg 1$ in the macroscopic limit $m_{\alpha}\rightarrow \infty$ and $\epsilon\rightarrow 0$. 

When $ m_{\alpha}\gg1$ and $\epsilon\ll1$, the leading-order terms in the asymptotic expansion of $\delta P_{L;1}$ are
\begin{equation}
\delta P_{L;1} = \frac{2}{1+ \epsilon m_{\alpha}}  \Big( 1+ \frac{\epsilon^2 m_{\alpha}}{(1+ \epsilon m_{\alpha})^2}  + \cdots \Big)
\label{eq:seventeen}.
\end{equation}
Finite-size effects at a given relative light intensity $\epsilon m_{\alpha}$ are encoded in the leading-power terms in $\epsilon \propto V^{-1}$, which may be explored in a cavity experiment.
Moreover, taking the macroscopic limit, we find 
\begin{equation}
\delta P_{L;1} = \frac{2}{1+ \epsilon m_{\alpha}} 
\label{eq:seventeen-2},
\end{equation}
which agrees in the form with the result for a Fock state. This is hardly surprising, however, as the relative uncertainty in photon number of a coherent state vanishes macroscopically, so that the statistical quantities are well approximated by their mean values.

\subsubsection{\label{sec:thermal}Thermal state}

Last, we consider a thermal state of photons interacting with the electron. A thermal state is a mixed quantum state, described by $\rho_{\textrm{ph}}(0) = \sum_{n=0}^{\infty}p_{\beta,n} \ket{n} \bra{n}$, where 
\begin{equation}
p_{\beta,n}= \frac{(m_{\beta})^n}{(m_{\beta}+1)^{n+1}} 
 \label{eq:eighteen},
\end{equation}
with $m_{\beta}=1/(e^{\beta \omega} -1)$ the mean photon number and $\beta=1/k_B T$ the inverse temperature. Unlike a coherent state, a thermal state hosts a large amount of thermal fluctuation, and the relative uncertainty in photon number does not vanish in the macroscopic limit. The initial and time-dependent total density matrices are $\rho_{\textrm{tot}}(0) = \sum_{n=0}^{\infty}p_{\beta,n} \ket{L,n} \bra{L,n}$ and $\rho_{\textrm{tot}}(t) = \sum_{n=0}^{\infty}p_{\beta,n} \ket{\Psi_n(t)} \bra{\Psi_n(t)}$, respectively, where $\ket{\Psi_n(t)}$ is given in Eqs.~(\ref{eq:three})(\ref{eq:six}). 

Following the previous procedures, we perform time average on the density matrix elements and obtain the leaked population 
$\delta P_L=\delta P_{L;1}-\delta P_{L;2}$, with
\begin{align}
\delta P_{L;1} & = \sum_{n=0}^{\infty} p_{\beta,n}  \frac{2}{1+\epsilon n}, \nonumber \\
\delta P_{L;2} & = \sum_{n=0}^{\infty} p_{\beta,n} \frac{3/2}{(1+\epsilon n)^2} 
\label{eq:nineteen}.
\end{align}
The expressions of $\delta P_{L;1}$ and $\delta P_{L;2}$ as functions of $\epsilon$ and $m_{\beta}$ are derived in Appendix~C of the Supplementary Materials. At large photon numbers, $\delta P_{L;2}$ is negligible.

We study $\delta P_{L;1}$ in the macroscopic limit $m_{\beta}\rightarrow \infty$ and $\epsilon\rightarrow 0$, demanding that $\epsilon m_{\beta}>1$ to avoid obtaining unphysical probabilities that are either negative or greater than one. When $\epsilon m_{\beta} \ll1$, one must include both $\delta P_{L;1}$ and $\delta P_{L;2}$ for a correct estimation of the degree of suppressed quantum tunneling. The lowest-order terms in the expansion of $\delta P_{L;1} $ are
\begin{equation}
\delta P_{L;1} = \frac{2}{\epsilon m_{\beta}}   \Big( \ln \epsilon m_{\beta}-\gamma_0 + \cdots \Big)
\label{eq:twenty},
\end{equation}
where $\gamma_0$ is Euler's constant.

A major difference between a thermal state and a coherent state (or a Fock state) is that the suppression of quantum tunneling is weakened in the former case due to the presence of thermal fluctuation introduced by photons partly destroying the quantum coherence underlying localization of the electron, reflected by an extra logarithm factor in the leaked population, Eq.~(\ref{eq:twenty}). This weakening effect originates from an incoherent superposition of different photon-number components in a thermal state and is \emph{not} damping in nature. In very high light, $\delta P_{L;1} \rightarrow 0$ because $1/x$ decays faster than $\ln x$ grows, and the electron will be permanently localized in $\ket{L}$. 

\begin{figure}
\centering
\includegraphics[width=3in]{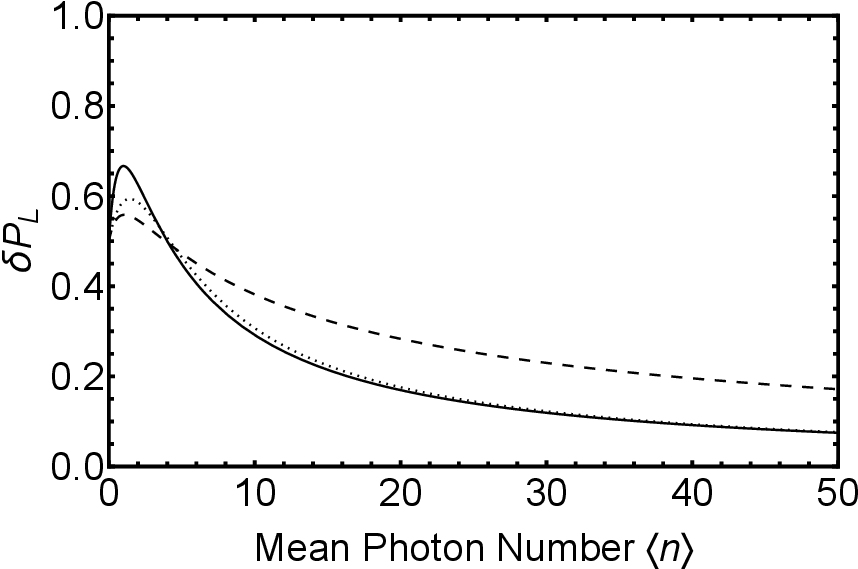}
\caption{Exact results of leaked population outside $\ket{L}$ of an electron in an isolated system interacting with a Fock state (solid curve), a thermal state (dashed curve) and a coherent state (dotted curve) of photons, as functions of (mean) photon numbers $\expect{n}=n$, $m_{\beta}$ and $m_{\alpha}$, respectively. The parameters are $\Delta_1 =\Delta =0$ and $\epsilon =0.5$.}
\label{fig:two}
\end{figure}

In Fig.~\ref{fig:two}, we compare exact results of $\delta P_{L}$ for a Fock state, a coherent state and a thermal state, given in Eqs.~(\ref{eq:eight}) (\ref{eq:sixteen})(\ref{eq:nineteen}), as functions of (mean) photon numbers $\expect{n}=n$, $m_{\alpha}$ and $m_{\beta}$, respectively. We have chosen a finite $\epsilon$ in order to distinguish between a Fock state and a coherent state, which would be identical in the macroscopic limit. At large photon numbers, $\delta P_{L}$ for a coherent state and for a Fock state become hardly discernible, whereas that for a thermal state diminishes at a much slower rate. In addition, we notice the existence of maxima at some finite photon numbers before $\delta P_{L}$ decays asymptotically. Mathematically, $\delta P_{L;2}$ are of comparable orders of magnitudes as (but still smaller than) $\delta P_{L;1}$ near these maxima. Physically, the electron is least localized in $\ket{L}$. Thus, upon increasing the intensity of light from zero, an electron in $\ket{L}$ is first leaked out to $\ket{R}$ and $\ket{R^*}$, and later trapped back in $\ket{L}$. A further numerical analysis (not shown) confirms the generality of this finding and that the locations of maxima move towards zero photons at larger $\epsilon$ settings. Variation of $\epsilon$ at a given photon number exhibits a similar behavior of $\delta P_{L}$. The maxima move closer to $\epsilon=0$ when a larger photon number is chosen. This is not surprising since $\epsilon$ and photon number enter the expressions of populations as a symmetric product. To first order, they should affect the populations in a similar fashion. 

\subsubsection{\label{sec:entropy}Linear entropy}

To see the direct effect of thermal fluctuation in photons on the electronic state, we numerically calculate linear entropy
 \begin{equation}
S(t) = 1- \text{Tr}\rho^2(t)
\label{eq:twenty-one},
\end{equation}
which measures the degree of mixture in a quantum state, and vanishes in a pure state, \emph{e.g.}, for a Fock state of photons. To this end, we must truncate the infinite summation in the photon density matrix. For a thermal state, for instance, we replace $\rho_{\textrm{ph}}(0) = \sum_{n=0}^{\infty}p_{\beta,n} \ket{n} \bra{n}$ with
 \begin{align}
\rho_{\textrm{ph}}^a(0) &= \sum_{n=0}^{N}p_{\beta,n} \ket{n} \bra{n},  \nonumber \\
\rho_{\textrm{ph}}^b(0) &=  \sum_{n=0}^{N-1}p_{\beta,n} \ket{n} \bra{n} + \Big(1- \sum_{n=0}^{N-1}p_{\beta,n}\Big)  \ket{N} \bra{N}
\label{eq:twenty-two},
\end{align}
and determine a choice of $N$ that yields negligible differences in the numerical results obtained using $\rho_{\textrm{ph}}^a(0)$ and $\rho_{\textrm{ph}}^b(0)$. We then use either method of truncation for the calculation.

\begin{figure}[ht]
\centering
\includegraphics[width=3in]{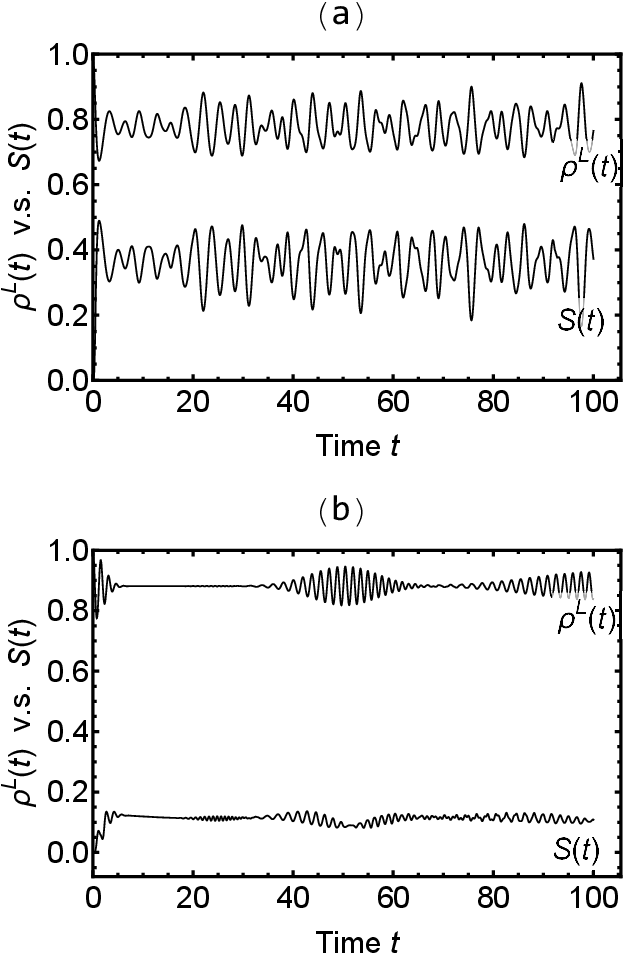}
\caption{Time-dependent linear entropy and population in $\ket{L}$ of an electron interacting with (a) a thermal state and (b) a coherent state of 16 photons in an isolated system. We observe an anti-correlation between the linear entropy and the population in (a), and collapses and revivals of Rabi oscillation with shortened time intervals in (b). The parameters are $\Delta_1 =\Delta =0$ and $g=\Gamma=1$.}
\label{fig:three}
\end{figure}

In general, we find that a thermal state of photons delivers more mixture to the electron than a coherent state of photons, indicated by a larger linear entropy. Moreover, the linear entropy is smaller for fewer incident photons, where the electron is less contaminated, reaches a maximum at an intermediate photon number, and diminishes towards zero at larger photon numbers where the electronic state becomes purer again as photons drive it towards an eigenstate of the Hamiltonian. In a thermal state, but not in a coherent state, we find an increasingly exact anti-correlation between the linear entropy and the population in $\ket{L}$ when the photon number is large enough, an example shown in Fig.~\ref{fig:three}(a), independent of the choice of parameters such as the detuning. In a coherent state, we observe collapses and revivals of Rabi oscillation of the population, shown in Fig.~\ref{fig:three}(b), with shortened intervals in time~\cite{JC-shore}. 
A possible explanation of the entropy-population anti-correlation is that the fluctuations in a thermal photon mode wash out much of the quantum interference effects between different levels. As a result, the purity of the electronic state is directly and solely determined by the overlap of the electron with the quasi-eigenstate $\ket{L}$, rather than by the complete quantum data stored in the wave function as a coherent superposition of $\ket{L}$, $\ket{R}$ and $\ket{R^*}$, as is the case when the electron interacts with a coherent photon mode.

\subsection{\label{sec:open}Open system: suppressed relaxation kinetics}

We now allow interaction of the three-level system with an environment via the excited level $\ket{R^*}$ and discuss the relaxation kinetics of an electron initially in $\ket{L}$ into the environment. We study this open system as a minimal model of the proposed bottleneck in photosynthetic transport, instead of invoking a more formal treatment which requires knowledge of microscopic details up and down the electron transport chain. We identify $\ket{L}$ with the immediate donors to P$700^+$ and P$680^+$. A phenomenological parameter $\gamma$ is used to characterize the rate of irreversible relaxation  from $\ket{R^*}$ to outside the system in the dark. In general, various processes may contribute to $\gamma$. Here we focus on transfers of photoexcited electrons from P$700^*$ and P$680^*$ to their immediate acceptors. Formally, $\gamma$ may be calculated by the resolvent method if interaction details with the environment is known.

For simplicity, we assume $\Delta_1=\Delta=0$ in the analytical treatment below, although the conclusion is not dependent on the convenient choice of parameters. We will first work with a Fock state of $n$ photons interacting with the three-level system. Cases of a coherent state and a thermal state of photons are discussed later.  

The Hamiltonian of the open system is $H'=H-(i\gamma/2)\ket{R^*}\bra{R^*}$. We solve the eigenvalue problem by perturbation theory, assuming $\gamma/\lambda_n<1$ for the convergence of the perturbative series. We define orthogonality of the eigenvectors of non-Hermitian operator $H'$ using eigenvectors of its hermitian conjugate $H'^{\dagger}$,~\cite{non-hermitian} obtained by sending $\gamma \rightarrow -\gamma$ in $H'$. 

\begin{figure}[ht]
\centering
\includegraphics[width=3in]{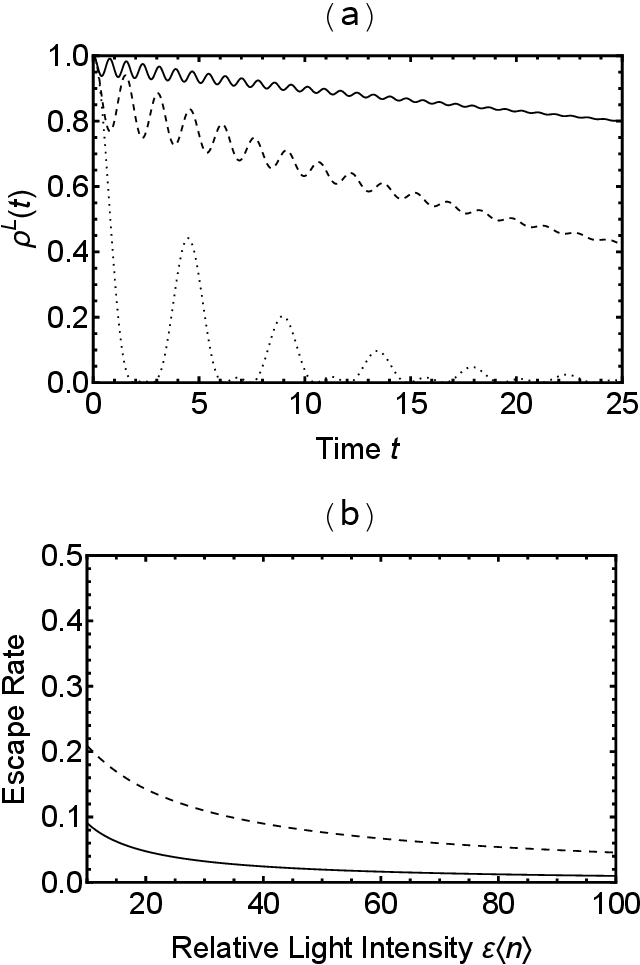}
\caption{(a) Time-dependent population in $\ket{L}$ of an electron interacting with a Fock state of $n=64$ (solid curve), $n=16$ (dashed curve) and $n=1$ (dotted curve) photons in an open system, where $\Delta_1 =\Delta =0$, $g=\Gamma=1$ and $\gamma=0.5$. (b) Escape rates (assumed to be proportional to photochemical rate constants in our interpretation) in unit of $\gamma$ of an electron interacting with a coherent (Fock) state (solid curve) and a thermal state (dashed curve) of photons, as functions of the relative light intensity $\epsilon\expect{n}$, plotted for $\epsilon\expect{n}> 10$.}
\label{fig:four}
\end{figure}

The eigenvectors of $H'$ are found to be
\begin{align}
\ket{0,n}' &=\ket{0,n} - i\gamma  \frac{\sin{\theta_n}\cos{\theta_n}}{2\sqrt{2} \lambda_n} \Big(\ket{+,n} - \ket{-,n} \Big), \nonumber \\
\ket{+,n}' &=\ket{+,n} + i\gamma \Big( \frac{\sin{\theta_n}\cos{\theta_n}}{2\sqrt{2} \lambda_n}  \ket{0,n}- \frac{\cos^2{\theta_n}}{8 \lambda_n} \ket{-,n} \Big),  \nonumber \\
\ket{-,n}' &=\ket{-,n}- i\gamma \Big( \frac{\sin{\theta_n}\cos{\theta_n}}{2\sqrt{2} \lambda_n}  \ket{0,n}- \frac{\cos^2{\theta_n}}{8 \lambda_n}  \ket{+,n} \Big)
\label{eq:twenty-three},
\end{align}
with eigenvalues $- i\kappa_0$, $\lambda_n-i\kappa$, $-\lambda_n-i\kappa$, respectively, where 
\begin{align}
\kappa_0 &=   \frac{\gamma}{2}\sin^2{\theta_n} , \nonumber \\
\kappa &=\frac{\gamma}{4}\cos^2{\theta_n}
\label{eq:twenty-four},
\end{align}
and $\ket{0,n}$, $\ket{+,n}$ and $\ket{-,n}$ are the unperturbed eigenvectors in Eq.~(\ref{eq:four}). 
We use these results to obtain time evolution of the system. Imagine in the distant past, the relaxation term was turned on adiabatically, and the eigenvectors gradually evolved into the ones in Eq.~(\ref{eq:twenty-three}).
At $t=0$, an electron is fed in $\ket{L}$ with forbidden backward hopping. The total quantum state is described by Eq.~(\ref{eq:five}) upon the replacements $\ket{0,n} \rightarrow \ket{0,n}'$, $\ket{+,n} \rightarrow \ket{+,n}'$ and $\ket{-,n} \rightarrow \ket{-,n}'$. The system then evolves according to the perturbed Hamiltonian $H'$. Because the perturbed eigenvectors do not coalesce within the parameter range of interest (\emph{i.e.}, no exceptional points exist), we can find time-dependent total wave function, as well as the diagonal elements of the reduced electron density matrix, using faithful eigenvector-decomposition of the initial wave function. 
In Fig.~\ref{fig:four}(a), we plot the probability that the electron remains in the initial state $\ket{L}$ at time $t$. It is clear that electron transport out of $\ket{L}$, and consequently from the system to the environment, is increasingly suppressed at larger photon numbers. This tendency is independent of the choice of $\gamma$, as long as the perturbation theory is well-defined, and persists in cases where $\Delta_1\neq0$ and $\Delta\neq0$ (not shown).

To proceed, we average out the fast oscillations in $\rho^{L}(t)$, $\rho^{R}(t)$, $\rho^{R^*}(t)$ as before and obtain
 \begin{align}
P_L(t)=&\cos^4{\theta_n}e^{- 2\kappa_0 t}+ \frac{1}{2} \sin^4{\theta_n}e^{- 2\kappa t}, \nonumber \\
P_R(t) = &\frac{1}{2}\sin^2{\theta_n}e^{-2 \kappa t}, \nonumber \\
P_{R^*} (t)=& \sin^2{\theta_n}\cos^2{\theta_n} \Big(e^{- 2\kappa_0 t} + \frac{1}{2}e^{- 2\kappa t}\Big)
\label{eq:twenty-five},
 \end{align}
respectively, where we have dropped the terms quadratic in $\gamma/\lambda_n$. We see that $P_L(t)$, $P_R(t)$ and $P_{R^*}(t)$ are all macroscopic quantities depending only on the relative light intensity $\epsilon n$ and the product $\gamma t$.

Using $\sin^2{\theta_n}=1/(1+\epsilon n)$, we find $P_R(t),P_{R^*}(t)\ll 1$ for $\epsilon n \gg1$. That is, in steady, high light illumination, a significantly larger amount of charge accumulates in $\ket{L}$, compared to that in low light. In addition, Eq.~(\ref{eq:twenty-five}) tells that when the incident light is strong enough, the escape rate of an electron from the system is inversely proportional to the light intensity. To see this, notice that the probability that the electron leaves the system at time $t$ is given by $1-P_L(t)-P_R(t)-P_{R^*}(t)$, which for $\epsilon n\gg1$ is well approximated by the leaked population $\delta P_L(t) = 1-P_L(t)$ defined previously, 
 \begin{equation}
\delta P_L (t) \simeq 1-\cos^4{\theta_n}e^{- 2\kappa_0 t}
\label{eq:twenty-six}.
\end{equation}
We identify $2\kappa_0 =  \gamma/(1+\epsilon n) $, the rate at which the electron is leaked out of $\ket{L}$, with the escape rate of the electron from the system which in our interpretation is directly representative of the rate constant of photochemistry. Physically, this means that photosynthetic efficiencies at PSI and at PSII are decided by the light-dependent reduction rates of P$700^+$ and P$680^+$, by electron transfers from PC and from TyrZ, respectively.

We now argue that the above results obtained for a Fock state of photons apply also to a coherent state and a thermal state of photons. For a coherent state, the relative variation $\delta (\epsilon n)/\epsilon m_{\alpha} = \delta n /m_{\alpha}$ vanishes as $m_{\alpha}\rightarrow \infty$, using the statistical property of the Poisson distribution in Eq.~(\ref{eq:fourteen}). Thus, Eqs.~(\ref{eq:twenty-five})(\ref{eq:twenty-six}) with $\epsilon n$ replaced by $\epsilon m_{\alpha}$ should be considered a good approximation in the macroscopic limit, in all light conditions. For a thermal state, we notice that the exponential function in  Eq.~(\ref{eq:twenty-six}) may be expanded in high light $\epsilon n\gg1$, to obtain $\delta P_L (t)  \simeq  (2 +\gamma t)/(1+\epsilon n)$ to leading order, where the only dependence on photon number is through the denominator. We can then perform a summation over photon number just as we did when studying an isolated three-level system (see Appendix~C in the Supplementary Materials), to go from the results for a Fock state to those for a thermal state. From previous discussion, we know that the summation produces an additional logarithm factor $\ln \epsilon m_{\beta}$ reflecting weakening of the suppression effect.  
The escape rates thus calculated are macroscopic quantities as functions of the relative light intensity $\epsilon\expect{n}$ and the relaxation parameter $\gamma$. As such, a Fock state and a coherent state are indistinguishable. Recall that $\gamma$ characterizes charge transfer rate from P$680^*$ (P$700^*$) to its immediate acceptor in complete darkness. One thus may compare the suppression effects of different photonic states as functions of light intensity on single electron transfer efficiency, by plotting the escape rates in unit of $\gamma$ to quantify the degree of suppression. This is shown in Fig.~\ref{fig:four}(b) for $\epsilon\expect{n}> 10$. Such a comparison is not possible in low light, where the reaction depends on more than one rate constant as can be seen from Eq.~(\ref{eq:twenty-five}).

In our simplified picture of photosynthetic transport, electrons are fed in $\ket{L}$ by LEF once and only if the system becomes unpopulated. In reality, this corresponds to fast binding/unbinding of PC to PSI, assumed to occur at a rate unaffected by changing light intensity. 
The only light dependence of local electron turnover is in the escape rate. When the escape rate is suppressed in higher light, an electron is trapped longer in the system with its population distribution given by Eq.~(\ref{eq:seven}). Since two electrons cannot coexist in the system, the rate at which the next electron is let in, commensurate with the escape rate, should also be lower. The picture may be compared to a bathtub with a constant water level, operated at equal injection and drainage rates which are tuned simultaneously.

In above discussion we have neglected radiative relaxation of the electron within the system, corresponding to prompt signals in a fluorescence measurement. An event of this type is incoherent and irreversible, and thus the electron will lose all the ``memories'' about previous trajectories. The electron then behaves no differently from an electron in $\ket{R}$ at the beginning of time evolution, which can be excited independently. At increasing intensities of light, independent photoexcitation gradually concedes to quantum correlated processes, which dynamically control the populations in the three levels, as we discuss in more detail in the next section.

\section{\label{sec:discussion}Discussion of Experiments}

In the traditional view of photosynthetic charge transfer, forward hops of electrons are a series of independent, incoherent and irreversible events. Passing of a photoexcited electron to the immediate acceptor is fast enough that it does not constitute a bottleneck on LEF~\cite{krause1991}. The rate-limiting step is thus believed to be the oxidation of plastoquinol at Cyt~$b6f$~\cite{haehnel1984}. When quantum correlations between charge donors and acceptors at photoexcitation are taken into account, however, this picture is challenged. As we have seen, when incident light is strong, the photochemical rate constant $\kappa_{\textrm{P}}$ changes inversely proportional to the intensity of light,
\begin{equation}
\kappa_{\textrm{P}} \propto  \frac{\gamma}{1+\epsilon \expect{n}}
\label{eq:twenty-seven},
\end{equation}
neglecting thermal fluctuation.
Quantum effects may suppress charge transfers from PC to P$700$ and from TyrZ to P$680$ so greatly that these steps become overall rate-limiting. This blockade effect is a direct action of light and may be termed photo-blockade.

The limiting nature of charge transfer from PC was already mentioned in early studies of transport in PSI. Under continuous illumination, it was found that partial inhibition of PC led to immediate decline in electron transport rate at PSI (ETR), contradicting the belief that LEF was controlled by slow plastoquinone turnover~\cite{haehnel1982}. Inhibition of charge feeding from PSII did not change the results. One may imagine that the illumination drove the system into the regime where Eq.~(\ref{eq:twenty-seven}) took control of the charge transfer dynamics. Application of PC inhibitors reduced in time the statistical average of the transition amplitude $\Gamma$ between PC and P$700$ (or equivalently raising $\epsilon$), and consequently the photosynthetic efficiency.

Our model allows for a qualitative description of the light responses of the PSI parameters Y(I), Y(NA) and Y(ND),~\cite{klughammer1994-PSI} defined as the ratios of photo-redox active (oxidizable) P700, non-oxidizable P700 and P$700^+$ to total P700, and commonly interpreted as the quantum yield of PSI, the acceptor-side limitation and the donor-side limitation on LEF, respectively. Previous study has maintained that shortage of PSI acceptors hardly occurs in steady state unless extreme environmental stresses are imposed,~\cite{baker2007} casting doubt on the common interpretation of Y(NA).
Applying the picture of photo-blockade, Y(I), Y(NA) and Y(ND) are instead correlated quantities that can be approximated, respectively, by the steady-state populations $P_{R}^{\textrm{ss}}$, $P_{R^*}^{\textrm{ss}}$ and $P_{L}^{\textrm{ss}}$ obtained by setting $t=0$ in Eq.~(\ref{eq:twenty-five}). We find 
\begin{align}
\text{Y(ND)} &\propto \Big( \frac{\epsilon\expect{n}}{1+\epsilon \expect{n}} \Big)^2  + \frac{1}{2} \Big( \frac{1}{1+\epsilon \expect{n}} \Big)^2, \nonumber\\
\text{Y(I)} &\propto \frac{1}{1+\epsilon \expect{n}}, \nonumber\\
\text{Y(NA)} &\propto  \frac{\epsilon\expect{n}}{(1+\epsilon \expect{n})^2}
\label{eq:twenty-eight}.
\end{align}
Here Y(NA) is understood to reflect a direct regulation by light of the population of P$700^*$ on the donor side.

The photo-blockade picture breaks down in very low light or at the onset of illumination, when there is a large number of ground-state P700 available. In this case, photoexcitation is little affected by charge transfer from PC, and the restriction on Y(NA) is absent. 
Moving from extremely low light to moderately low light, charge gradually builds up in the PC pool, while photo-blockade becomes increasingly pronounced and then dominant. Eventually the system enters a light-driven steady state in which the population of oxidizable P700 is entirely determined by the reduction rate of P$700^+$. In this light, photo-blockade plays the role of a protective barrier against excess charge excitations. 
In extremely high light, or in the absence of $\Delta$pH-inducing proteins such as PGR5 or PGRL1, insufficient dissipation of excitation energy may cause overheating and destruction of any quantum coherence. Without the protection of photo-blockade on the donor side, charge excitations flood the acceptor side of PSI and lead to unregulated charge recombination~\cite{ph-pgr5}. This is signaled by an anomalous rise in Y(NA),~\cite{zivcak2015,huang2010} which now reflects the capacity of the acceptor side to accommodate excitations.

\begin{figure}[ht]
\centering
\includegraphics[width=3in]{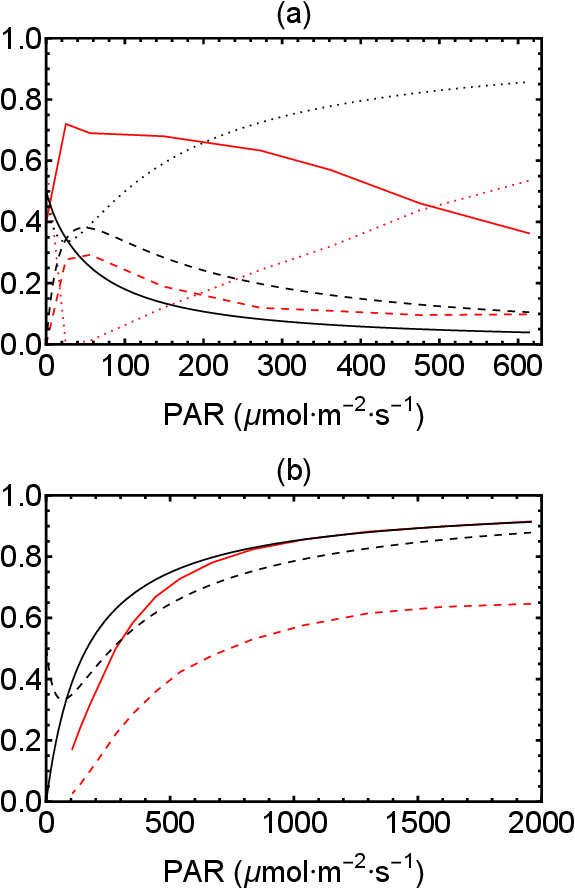}
\caption{(a) Comparison of steady-state populations $P_{R}^{\textrm{ss}}$ (black solid curve), $P_{R^*}^{\textrm{ss}}$ (black dashed curve) and $P_{L}^{\textrm{ss}}$ (black dotted curve) with measured light response data of Y(I) (red solid curve), Y(NA) (red dashed curve) and Y(ND) (red dotted curve), respectively, adapted from Ref.~\cite{klughammer2008-PSI}. 
 (b) Comparison of theoretically calculated $\text{ETR}/\text{ETR}_{\text{max}}$ (black solid curve) and $P_{L}^{\textrm{ss}}$ (black dashed curve), both approaching unity at infinite light intensity, with measured moderate-to-high-light behaviors of ETR (red solid curve) and Y(ND) (red dashed curve), respectively, adapted from Ref.~\cite{zivcak2015}. The amplitude of measured ETR has been scaled to match the high light asymptote of calculated $\text{ETR}/\text{ETR}_{\text{max}}$. 
The scaling parameter $b$ takes different values in (a) and in (b), as one would expect for measurements in different samples.}
\label{fig:six}
\end{figure}

In Fig.~6(a) we compare $P_{R}^{\textrm{ss}}$, $P_{R^*}^{\textrm{ss}}$ and $P_{L}^{\textrm{ss}}$ with a set of experimentally obtained light response data of Y(I), Y(NA) and Y(ND), respectively~\cite{klughammer2008-PSI}. 
The proportionality factor $b$ between the relative light intensity $\epsilon\expect{n}$ and photosynthetically active radiation (PAR) measured in the lab (denoted by $I_{\text{PAR}}$), $\epsilon\expect{n} = b I_{\text{PAR}}$, can be obtained by concurrently scaling the horizontal axes of theoretical predictions to match the data. With a proper choice of $b$, matching of the minima of $P_{L}^{\textrm{ss}}$ and Y(ND) and matching of the maxima of $P_{R^*}^{\textrm{ss}}$ and Y(NA) can be simultaneously achieved, and theory seems to agree with data in general trend, with discrepancies that may have been caused by thermal fluctuation in light. As we have seen in Fig.~\ref{fig:two}, thermal fluctuation tends to drive down Y(ND). 
Other deviations may be due to the insufficiency in assuming a conserved local charge, apart from the microscopic details not addressed in our model and unknown factors in the experiment.
Especially in low light while PSI transits from an equilibrium state to a light-driven steady state, the rapid rise in LEF may be accompanied with an increasing average local charge, possibly due to accelerated binding dynamics of PC to PSI. A more reduced pool of P$700$ implies gains in the numerators of Y(I) and Y(NA), and a loss in the numerator of Y(ND), while the common denominator of the three parameters, the total number of P700 molecules, remains a constant. Using Y(I) + Y(NA) + Y(ND) =~1, the total gain in Y(I) and Y(NA) equals the  loss in Y(ND). 
A further insight comes from the ratio
\begin{equation}
\frac{P_{R^*}^{\textrm{ss}}}{P_{R}^{\textrm{ss}}} \propto \frac{\epsilon\expect{n}}{1+\epsilon \expect{n}},
\label{eq:twenty-nine}
 \end{equation}
which shows that in low light electrons tend to dwell in $\ket{R}$ rather than in $\ket{R^*}$, and consequently a substantial portion of the loss in Y(ND) goes to Y(I) rather than to Y(NA). That is, the deviations of Y(ND) and Y(I) from $P_{L}^{\textrm{ss}}$ and $P_{R}^{\textrm{ss}}$, respectively, due to an increasing (rather than conserved) local charge compensate each other, while Y(NA) stays roughly unaffected until higher light where it starts to share the gain with Y(I). This explains the rise of Y(I) in low light, in contrast to the monotonic decline of $P_{R}^{\textrm{ss}}$. It also explains the much deeper dip of Y(ND) than $P_{L}^{\textrm{ss}}$.  

The discrepancy between $\kappa_{\textrm{P}}$ and measured ETR is also apparent. This is resolved by adopting a picture of “multi-channel” transport. While the transfer efficiency of a single electron subject to photo-blockade is described by $\kappa_{\textrm{P}}$, the number of channels contributing to transport depends linearly on the exciton density, which in turn is proportional to the density $\epsilon \expect{n}$ of incident photons. Thus we may write $\text{ETR} =\kappa_{\textrm{P}}\epsilon \expect{n}$ in steady state, or
\begin{equation}
\text{ETR} = \frac{\epsilon\expect{n}}{1+\epsilon \expect{n}}  \text{ETR}_{\text{max}}
\label{eq:thirty},
\end{equation}
using Eq.~(\ref{eq:twenty-seven}), where $\text{ETR}_{\text{max}}$ is the ETR at infinite light intensity. By definition $\text{ETR}_{\text{max}}$ has absorbed the parameter $\gamma$. Eqs.~(\ref{eq:twenty-eight})(\ref{eq:thirty}) tell us that both ETR and Y(ND) approach constant values in high light, and that ETR grows faster than Y(ND). These predictions are compared with experiment in Fig.~6(b).  We find reasonable agreement of theoretical and measured ETR in moderate to high light, but a difference between the two exists in low light which may be due to the effects of structural factors of the sample~\cite{thornley1976}. Also as mentioned earlier, Eq.~(\ref{eq:twenty-seven}) is valid only when incident light is strong enough, and this may have contributed to the difference. On the other hand, thermal fluctuation may account for the difference between theoretical and measured Y(ND), among other factors.
In our picture, saturation of ETR in high light originates from photo-blockaded transport, while the notion of openness/closedness of a reaction center controlled by the redox state of the quinone acceptor is less needed. The operating efficiency of PSI, defined as ETR divided by irradiance, is proportional to $\kappa_{\textrm{P}}$. 

Fig.~6 shows that the low-light part of Y(NA) may be used to calibrate instrumentation before performing measurements. A continuous-wave laser with a narrow linewidth is preferable to approximate single-mode photons and to minimize the influence of thermal fluctuation. Calibration should be done under constant ambient conditions so that the first factor in Eq.~(\ref{eq:eleven}) can be treated as a constant and then light intensity is the only variable to work with. The proportionality factor $b$ thus extracted clarifies the definitions of ``high light'' and ``low light'' for a specific setup, and the data sets a reference for future measurements with different light sources and under different ambient conditions. In principle, one may construct the action spectrum of photo-blockade by adding up contributions from lights of all wavelengths.

\begin{figure}
\centering
\includegraphics[width=3in]{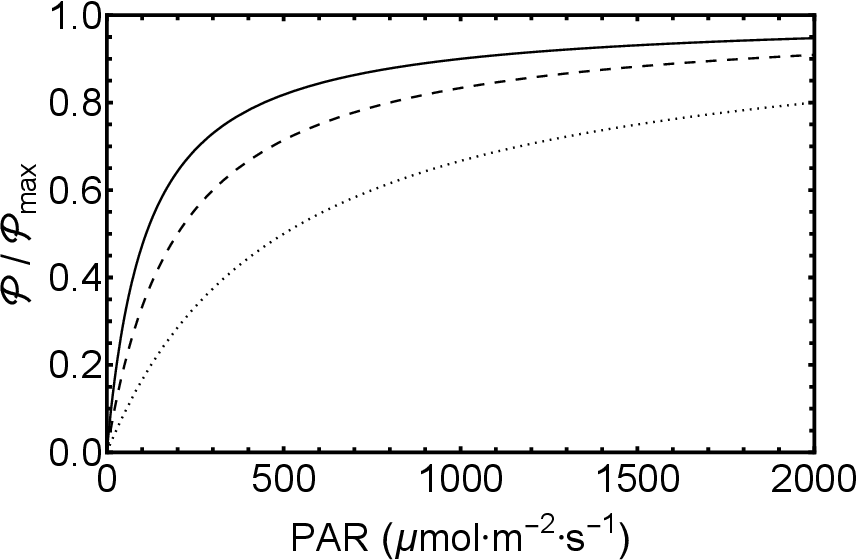}
\caption{Light response of normalized gross photosynthetic rate by Eq.~(\ref{eq:thirtyone}), for $b=0.009$ (solid curve), $0.005$ (dashed curve) and $0.002$ (dotted curve) in unit of $(\mu \text{mol} \cdot m^{-2}\cdot s^{-1})^{-1}$.}
\label{fig:seven}
\end{figure}

ETR in Eq.~(\ref{eq:thirty}) is of the Michaelis–Menten form. 
On the acceptor side of PSI, the transported electrons are utilized  to generate ATP for carbon fixation. In theory, four electrons are needed to fix one carbon dioxide molecule~\cite{long2003}. 
Assuming a linear relation between ETR and gross photosynthetic rate $\mathcal{P}$ (the rate of carbon fixation neglecting respiration), we find
\begin{align}
\mathcal{P}=  \frac{b  I_{\text{PAR}} }{1+b I_{\text{PAR}}} \mathcal{P}_{\text{max}}
\label{eq:thirtyone},
\end{align}
using $\epsilon \expect{n} = b I_{\text{PAR}}$, where $\mathcal{P}_{\text{max}}$ is the maximum gross photosynthetic rate. In reality, $\mathcal{P}_{\text{max}}$ depends on various ambient factors such as temperature and carbon dioxide concentration. 
The hyperbolic formula Eq.~(\ref{eq:thirtyone}) and its refinements have been successfully used to fit the light response data~\cite{thornley1976}. Here we attribute its physical origin to photo-blockade. The factor $b$ determines the convexity of the light response curve, as we show in Fig.~7. 
Using Eq.~(\ref{eq:eleven}), we have  
\begin{equation}
b \propto  \frac{ |\braopket{R}{\vec{\xi} \cdot \hat{\vec{d}}}{R^*}|^2}{\Gamma^2}
\label{eq:thirtytwo}.
\end{equation}
Thus, $b$ depends on light–matter coupling strength, an intrinsic property of the photosystem, and the transition amplitude $\Gamma$, which is expected to be temperature-dependent~\cite{marcus1985}. In general, experimentally obtained $b$ may also depend on leaf absorptance and supply of carbon dioxide in carboxylation.

Eqs.~(\ref{eq:twenty-seven})(\ref{eq:thirty}) may be used to understand the light responses of PSII parameters. To this end, again, one is reminded that the equations imply a donor-side scheme, whereas current interpretation of fluorescence measurements is based on the conceived limitation on the acceptor side,~\cite{krause1991,baker2008-PSII} where a reduced quinone acceptor is thought to produce a ``closed'' reaction center. While it is tempting to compare $\kappa_{\textrm{P}}$ with the fraction of ``open'' centers in a ``puddle model'' or a ``lake model'', and the light responses of both indeed bear a resemblance,~\cite{kramer-qL} it should be clear that these are quantities of distinct origins. In particular, $\kappa_{\textrm{P}}$ should be understood as a holistic parameter characterizing the operating efficiency of PSII, \emph{e.g.}, corresponding to in the acceptor-side scheme the combined limiting effects of the degree of openness of reaction centers and the maximum yield of an open reaction center. It is the usually measured quantum yield at PSII. Together $\gamma$ and $\epsilon$ (or equivalently, $b$) decide the shape of the light response curve of $\kappa_{\textrm{P}}$, while that of NPQ follows a compensatory pattern. The same is true for the curve of electron transport rate at PSII, which generally differs in convexity $b$ from its counterpart at PSI~\cite{zivcak2015}. We note that although the quantum yields at PSII and at PSI, $\kappa_{\textrm{P}}$ and Y(I), share the same light dependence by Eqs.~(\ref{eq:twenty-seven})(\ref{eq:twenty-eight}), their physical meanings are different. Last, a nonzero detuning $\Delta$ also affects the shape of a light response curve, as shown in Section~\ref{sec:model}.

The picture of photo-blockade was motivated also by data from photodamaged PSII. Once associated with over-reduced quinone acceptors facilitating formation of reactive oxygen species,~\cite{vass1992} photodamage is increasingly believed to take place at the oxygen-evolving complex. A two-step mechanism hypothesized that excitation of manganese ions by ultraviolet light impaired the oxygen-evolving complex and blocked charge transfer from TyrZ to P$680$~\cite{hakala2005,ohnishi2005,allakhverdiev2005}. Aggregation of highly oxidizing P$680^+$ then caused irreversible damage to the functional components. This donor-side mechanism explains why ultraviolet light is more efficient than visible light in damaging PSII,~\cite{hakala2005,ohnishi2005} and
is supported by evidence that the rate of photodamage is unaffected by  inhibition of transport on the acceptor side~\cite{allakhverdiev2005}. Nevertheless, it fails to account for the drastic increase in photodamaging efficiency when the wavelength of incident light approaches 680nm~\cite{jones1966,santabarbara2001}. Such a feature, more obvious \emph{in vitro} in the absence of the protection of leaves, is easily explained by photo-blockade, which is likewise a donor-side mechanism consistent with the fact that the rate of photodamage is directly proportional to the intensity of  light,~\cite{allakhverdiev2004,tyystjarvi1996} and whose effect is most prominent when  light is resonant with excitation of P$680$.
We suggest that long-lived P$680^+$ may be produced by the action of red light suppressing charge transfer from still healthy oxygen-evolving complex. This mechanism functions together with the aforementioned manganese mechanism highlighting detrimental effects of ultraviolet light on the oxygen-evolving complex. The  action spectrum of photodamage thus manifests a superposition of the effects of both mechanisms.

\section{\label{sec:conclusion}Concluding Remarks}

We have proposed a new dynamic mechanism to explain the modulation of photosynthetic efficiencies under varying light conditions, taking into consideration the quantum correlations between donor and acceptor molecules in electron transport. The declines in the quantum yields of PSI and PSII at increasing irradiance are attributed to light-induced blockade of charge transfers from PC to P700 and from TyrZ to P680, respectively.
Using a minimal model of three quantum levels, we derive a set of formulas that describe qualitatively the light response curves of photosynthetic parameters. The model also offers interpretations for the observed PC-restricted LEF and recent data from photodamaged PSII.

Our main findings can be summarized as follows. First, the occurrence of photo-blockade does not depend on the degree of coherence in incident light or the temporal correlations imparted to the system upon excitation~\cite{brumer1991}. In defining the quantity ``leaked population'' that measures the degree of photo-blockade, we average out oscillations in time evolution of the system. The results show that suppression of quantum tunneling and relaxation kinetics can be induced by both a coherent light and a thermal light, although in the latter case the effect is weakened by thermal fluctuation. Second, all photon modes in sunlight contribute to photo-blockade, the degree of which diminishes as the photon energy drifts away from resonance with the electronic transition. Third, photo-blockade is a relevant concept both in an optical cavity and in free space (infinitely large system size). Our results in free space bear physical significance and may be compared with photosynthetic measurements.

As a direct test of our model, we suggest measurement of the quantum yield of PSII with different light sources. Quantum mechanically, light emanating from a laser and from a filtered incandescent lamp may be approximated, respectively, by a coherent state and a thermal state of photons. For the measurement, illumination should be continuous, steady, and of a narrow linewidth.
An observable difference is expected in the tails of the light response curves measured with a lamp and with a laser. In the former case, thermal fluctuation weakens the photo-blockade effect and thus one may find a higher quantum yield at a given light intensity. The difference should be more pronounced in high light (\emph{e.g.}, PAR > $1000\mu \text{mol} \cdot m^{-2}\cdot s^{-1}$), where the dependences on light intensity of leaked population and of escape rate coincide, and the logarithm factor in Eq.~(\ref{eq:twenty}) may manifest itself in the data (see the discussion before Fig.~\ref{fig:four}(b)). In addition, one may observe an apparent growth of the quantum yield with temperature in the case of lamp illumination, because the mean photon number in a thermal state is positively correlated with temperature. This applies also to filtered sunlight, the coherence time of which was recently obtained experimentally~\cite{ricketti2022}. 
In principle, the same measurement may be conducted on PSI. However, PSI is more vulnerable than PSII to photodamage in high light, and the data may be obscured by feedback processes. We expect the measurement on PSII to be more revealing of the effect of thermal fluctuation.

On the other hand, evidence of an over-reduced PC pool in a light-driven steady state, \emph{e.g.}, a local charge imbalance between PC and P700 described by Eq.~(\ref{eq:twenty-five}) near $t=0$, may be obtained by measuring changes in the amplitude of the fastest component of P$700^+$ reduction kinetics induced by a saturating light pulse with an actinic light of varied intensities in the background, and comparing them with predictions in equilibrium. To this end, extra reduced PC may be added to ensure an ample supply of PC to PSI for the formation of charge transfer complexes,~\cite{haehnel1996} thus neutralizing the redistribution effects of (moderate) variations in the actinic light. We expect a local charge imbalance to be resolvable, given that the release of oxidized PC from the complex is much slower than the rebinding of next PC~\cite{finazzi2005}.

We emphasize that the analysis presented in this work is at the level of phenomenology. Further testimonies and refinements of our results are encouraged. It would be interesting to see how our predictions are compared with data in the lab. Also, it remains to be investigated whether photo-blockade would survive the incorporation of more realistic interactions. For instance, to study more natural environments and possible non-radiative pathways of relaxation (\emph{e.g.}, at conical intersections), one may allow coupling of the electronic states to a continuum of photonic and/or vibrational modes, and replace the three-level system with a realistic molecular system. 
On the other hand, thanks to the phenomenological treatment our results do enjoy some degree of robustness. For instance, experiments revealed that instead of acting as a source of dissipation, electronic-vibrational coupling more likely played a facilitating role in the forward transfer of electrons from excited special pair~\cite{fuller2014,romero2014}. In our modeling, this translates to an enhanced $\gamma$, yet Eqs.~(\ref{eq:twenty-eight})(\ref{eq:twenty-nine})(\ref{eq:thirty}) as well as the inverse proportionality of the photochemical rate constant $\kappa_{\textrm{P}}$ to light intensity remain unchanged.

Two simple extensions of our model are of particular interest. The first is to replace the quantized description of light with a classical one, and see if the blockade effect may be reproduced. With a semi-classical approach, the dependence of the blockade effect on the degree of coherence of light may be explored~\cite{brumer1991}. 
The second addresses possible incoherent transfer of excitons from the antenna to the reaction center by expanding the two levels $\ket{R},\ket{R^{*}}$ in Eq.~(\ref{eq:two}) to four levels $\ket{R_\text{RC}},\ket{R_{\text{RC}}^{*}},\ket{R_\text{A}},\ket{R_\text{A}^{*}}$, where $\ket{R_\text{RC}}$ and $\ket{R_{\text{RC}}^{*}}$ stand for the ground state and the excited state of P700 coupled strongly to light, respectively, and $\ket{R_\text{A}},\ket{R_\text{A}^{*}}$ describe all of the antenna as a unity. Here we give a heuristic argument of why photo-blockade may be found in this five-level system. In this case, light assumes two effects: to populate $\ket{R_\text{A}^{*}}$ with an exciton density proportional to the density of incidence photons, and to affect tunneling of every electron out of the initial state $\ket{L}$. Because transfer of an exciton from the antenna to the reaction center by F\"{o}rster-type pairwise hopping (Coulombic dipole–dipole interaction) necessarily involves synchronous excitation of an electron from $\ket{R_\text{RC}}$ to $\ket{R_{\text{RC}}^{*}}$ and de-excitation of $\ket{R_\text{A}^{*}}$ to $\ket{R_\text{A}}$, the prerequisite for it to happen is that P$700^+$ has been re-reduced, \emph{e.g.}, by an electron that tunneled from $\ket{L}$ to $\ket{R_\text{RC}}$. If we assume Eq.~(\ref{eq:twenty-six}) is somehow still useful, this electron tunneling will be suppressed as light gets stronger. So will thus be the overall photosynthetic charge transfer through PSI. A formal analysis of the five-level system is however needed to substantiate this argument.

Photo-blockade is a fundamental quantum effect that may find evidence beyond photosynthetic organisms. At the molecular level, suppressed electron tunneling over a prolonged period of time produces abnormal concentrations of certain reactants which may slow down biochemical processes and induce irreversible mutations, \emph{e.g.}, in a DNA sequence. In a larger scope and on a much longer time scale, regional and temporal patterns of solar-radiation related mutations may constitute a dynamic factor behind punctuated morphological evolution of biological species~\cite{gould1977}.

%\appendix

%\section{Analytic solution for degenerate $\ket{L}$ and $\ket{R}$}
%\section{Leaked population for a coherent state and a thermal state of photons}

%%       \cite{<label>} ==> [1]

%%




\section*{Acknowledgments}
We thank Terufumi Yamaguchi and Anton Kockum for valuable comments. This work was supported by JSPS Kakenhi Grant No. 21H01034.

\newpage

\appendix

% The following command adds "S" to reference numbers
\renewcommand{\bibnumfmt}[1]{[S#1]}
\renewcommand{\citenumfont}[1]{S#1}

\renewcommand{\thefigure}{A\arabic{figure}}
\setcounter{figure}{0}

\section{Population data complementary to Figure~2 of the main text}

\begin{figure}[!htb]
\centering
\includegraphics[width=3in]{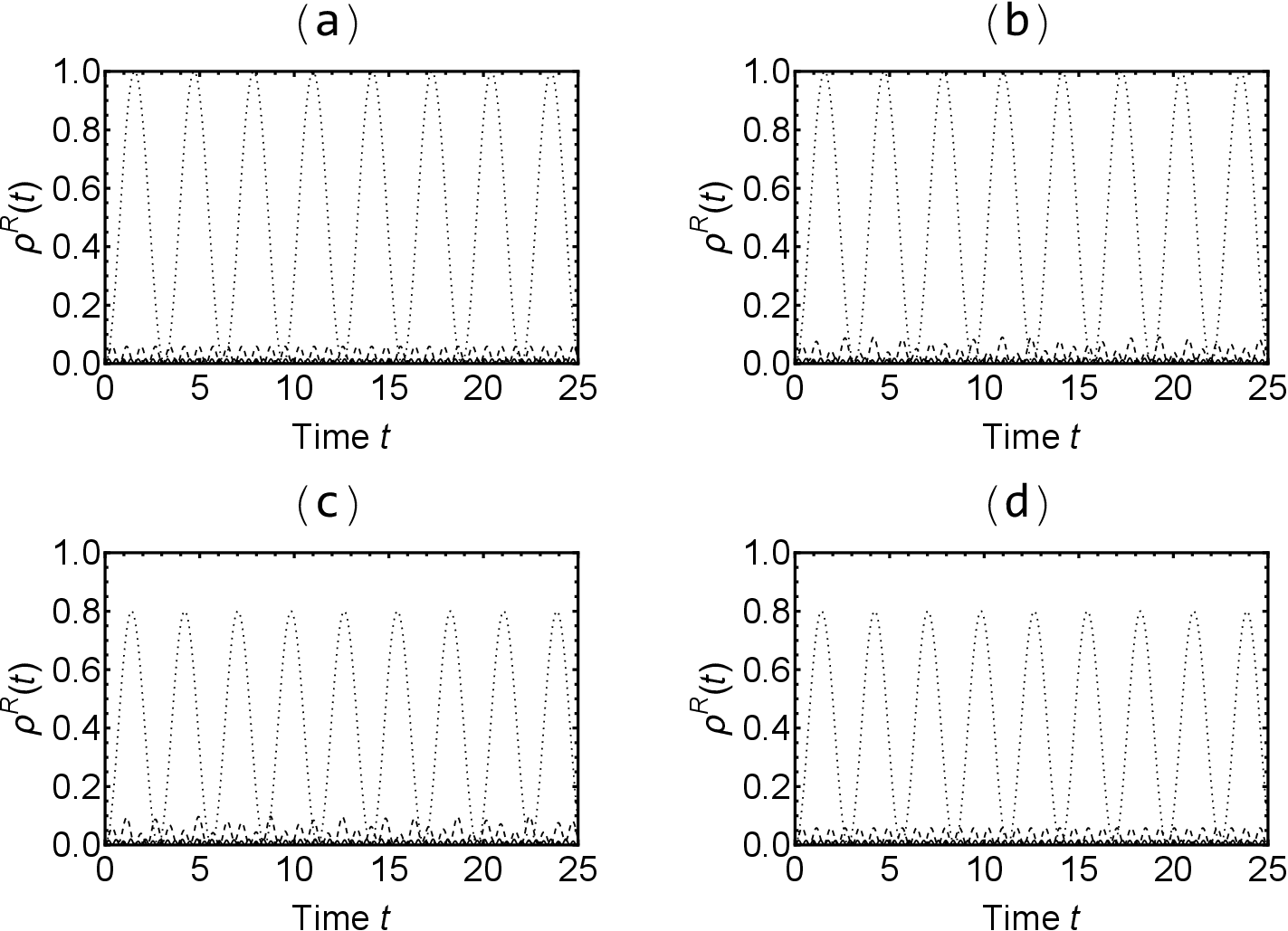}
\caption{Time-dependent population in $\ket{R}$ of an electron interacting with a Fock state of $n=64$ (solid curves), $n=16$ (dashed curves) and $n=0$ (dotted curves) photons in an isolated system. The parameters are $\Delta_1 =0$ and $\Delta =0$ in (a), $\Delta_1 =0$ and $\Delta =1$ in (b), $\Delta_1 =1$ and $\Delta =0$ in (c), and $\Delta_1 =1$ and $\Delta =1$ in (d). In all the plots $g=\Gamma=1$.}
\label{fig:appendixA-one}
\end{figure}

\begin{figure}[!htb]
\centering
\includegraphics[width=3in]{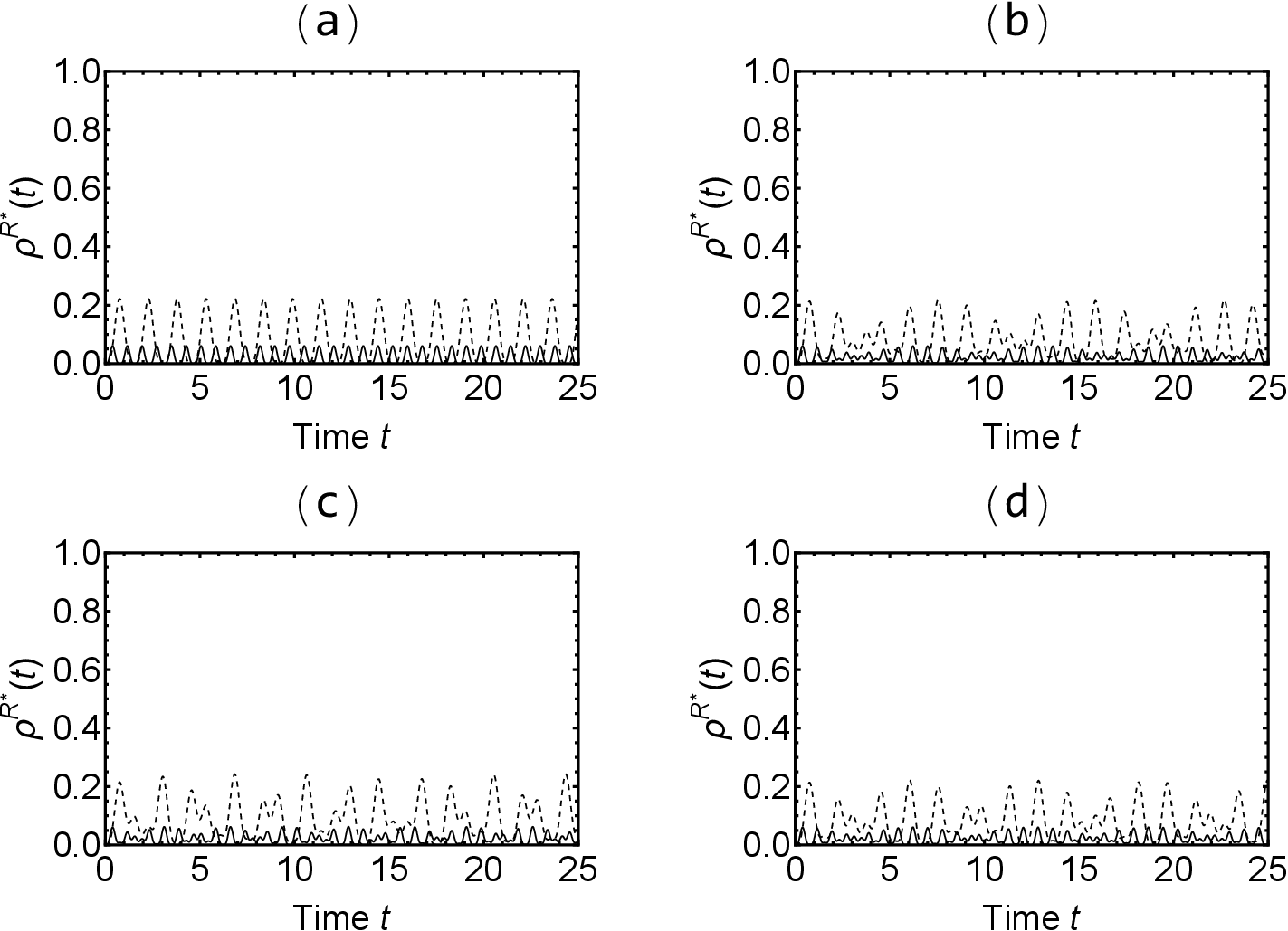}
\caption{Time-dependent population in $\ket{R^*}$ of an electron interacting with a Fock state of $n=64$ (solid curves) and $n=16$ (dashed curves) photons in an isolated system. At $n=0$ the state is unpopulated. Parameter settings in the subfigures are the same as in Fig.~\ref{fig:appendixA-one}.}
\label{fig:appendixA-two}
\end{figure}

\clearpage

\section{Analytic solution for degenerate $\ket{L}$ and $\ket{R}$}

In this appendix, we solve the time-dependent Schr\"{o}dinger equation analytically, assuming $\ket{L}$ and $\ket{R}$ are degenerate, $\Delta_1=0$. Substituting the wave function in Eq.~(\ref{eq:three}) of the main text, the Schr\"{o}dinger equation becomes
\begin{equation}
\left( \begin{array}{c}
\dot{c_{1,n}}\\
\dot{c_{2,n}}\\
\dot{c_{3,n}}
 \end{array} \right)= 
-i H_n \left( \begin{array}{c}
c_{1,n}\\
c_{2,n}\\
c_{3,n}
 \end{array} \right) 
 \label{appeq:one}.
 \end{equation}
 The solution takes the form
\begin{equation}
\left( \begin{array}{c}
c_{1,n}(t)\\
c_{2,n}(t)\\
c_{3,n}(t)
 \end{array} \right)= a (t) 
  \left( \begin{array}{c}
c_{1,n}(0)\\
c_{2,n}(0)\\
c_{3,n}(0)
 \end{array} \right)
  \label{appeq:two},
 \end{equation}
 with $c_{1,n}(0) = 1$, $c_{2,n}(0) = 0$, and $c_{3,n}(0) = 0$ given that the electron is initially in $\ket{L}$. 
Using orthogonality of the eigenvectors of $H_n$, one can show that $a (t) $ is a symmetric matrix. 
 
We solve Eq.~(\ref{appeq:one}) by Laplace transform and the theory of cubic equations. The elements of $a (t) $ are found to be,
\begin{strip}
\begin{align}
a_{11}&=\frac{\lambda_1^2-\Delta \lambda_1-g_n^2}{(\lambda_1-\lambda_2)(\lambda_1-\lambda_3)} e^{-i\lambda_1 t} + \frac{\lambda_2^2-\Delta \lambda_2-g_n^2}{(\lambda_2-\lambda_1)(\lambda_2-\lambda_3)} e^{-i\lambda_2 t} + \frac{\lambda_3^2-\Delta \lambda_3-g_n^2}{(\lambda_3-\lambda_1)(\lambda_3-\lambda_2)}e^{-i\lambda_3 t}, \nonumber \\
a_{12}&=\frac{(\lambda_1- \Delta)\Gamma}{(\lambda_1-\lambda_2)(\lambda_1-\lambda_3)} e^{-i\lambda_1 t} + \frac{(\lambda_2- \Delta)\Gamma}{(\lambda_2-\lambda_1)(\lambda_2-\lambda_3)} e^{-i\lambda_2 t} + \frac{(\lambda_3- \Delta)\Gamma}{(\lambda_3-\lambda_1)(\lambda_3-\lambda_2)}e^{-i\lambda_3 t}, \nonumber \\
a_{13}&=\frac{g_n \Gamma}{(\lambda_1-\lambda_2)(\lambda_1-\lambda_3)} e^{-i\lambda_1 t} + \frac{g_n \Gamma}{(\lambda_2-\lambda_1)(\lambda_2-\lambda_3)} e^{-i\lambda_2 t} + \frac{g_n \Gamma}{(\lambda_3-\lambda_1)(\lambda_3-\lambda_2)}e^{-i\lambda_3 t}, \nonumber \\
a_{22}&=\frac{(\lambda_1- \Delta)\lambda_1}{(\lambda_1-\lambda_2)(\lambda_1-\lambda_3)} e^{-i\lambda_1 t} + \frac{(\lambda_2- \Delta)\lambda_2}{(\lambda_2-\lambda_1)(\lambda_2-\lambda_3)} e^{-i\lambda_2 t} + \frac{(\lambda_3- \Delta)\lambda_3}{(\lambda_3-\lambda_1)(\lambda_3-\lambda_2)}e^{-i\lambda_3 t}, \nonumber \\
a_{23}&=\frac{g_n \lambda_1}{(\lambda_1-\lambda_2)(\lambda_1-\lambda_3)} e^{-i\lambda_1 t} + \frac{g_n \lambda_2}{(\lambda_2-\lambda_1)(\lambda_2-\lambda_3)} e^{-i\lambda_2 t} + \frac{g_n \lambda_3}{(\lambda_3-\lambda_1)(\lambda_3-\lambda_2)}e^{-i\lambda_3 t}, \nonumber \\
a_{33}&=\frac{\lambda_1^2- \Gamma^2}{(\lambda_1-\lambda_2)(\lambda_1-\lambda_3)} e^{-i\lambda_1 t} + \frac{\lambda_2^2- \Gamma^2}{(\lambda_2-\lambda_1)(\lambda_2-\lambda_3)} e^{-i\lambda_2 t} + \frac{\lambda_3^2- \Gamma^2}{(\lambda_3-\lambda_1)(\lambda_3-\lambda_2)}e^{-i\lambda_3 t}
  \label{appeq:three},
 \end{align}
 \end{strip}
where the real numbers
 \begin{align}
\lambda_{1}&=\frac{\Delta}{3} -2\sqrt{-P}\: \cos{\Big( \frac{1}{3}\arccos{\frac{Q}{2(-P)^{\frac{3}{2}}}}\Big)}, \nonumber \\
\lambda_{2}&=\frac{\Delta}{3} -2\sqrt{-P}\: \cos{\Big( \frac{1}{3}\arccos{\frac{Q}{2(-P)^{\frac{3}{2}}}}+\frac{2\pi}{3} \Big)}, \nonumber \\
\lambda_{3}&=\frac{\Delta}{3} -2\sqrt{-P}\: \cos{\Big( \frac{1}{3}\arccos{\frac{Q}{2(-P)^{\frac{3}{2}}}}-\frac{2\pi}{3} \Big)}
  \label{appeq:four}
 \end{align}
are eigenvalues of $H_n$, with 
 \begin{align}
 P&= -\frac{\Delta^2}{9}-\frac{\Gamma^2+g_n^2}{3}, \nonumber \\
Q&= -\frac{2\Delta^3}{27}-\frac{(g_n^2-2\Gamma^2)\Delta}{3}
  \label{appeq:five}.
 \end{align}
 At zero detuning $\Delta =0$, the expressions of the eigenvalues are greatly simplified,
  \begin{align}
\lambda_1&=-\lambda_n,   \nonumber\\
\lambda_2&=\lambda_n, \nonumber\\
\lambda_3&=0
  \label{appeq:six},
 \end{align}
 where $\lambda_n = \sqrt{\Gamma^2+g_n^2}$.

\section{Leaked population for a coherent state and a thermal state of photons}

In this appendix, we obtain compact expressions of leaked population $\delta P_L$ in an isolated system. For a coherent state of photons, we start with Eq.~(\ref{eq:sixteen}) in the main text. The summation in $\delta P_{L;1}$ was encountered in the study of linear response of a charged Bose gas immerse in an external magnetic field, and was known to produce a confluent hypergeometric function. The summation in $\delta P_{L;2}$ produces a generalized hypergeometric function. We find 
\begin{align}
\delta P_{L;1} & = 2 e^{- m_{\alpha}}  \: _1 F_1  \left(\frac{1}{\epsilon}; 1+\frac{1}{\epsilon} ; m_{\alpha} \right),  \nonumber \\
\delta P_{L;2}&=\frac{3}{2} e^{- m_{\alpha}}  \: {}_2 F_2  \left(\frac{1}{\epsilon}, \frac{1}{\epsilon}; 1+\frac{1}{\epsilon}, 1+\frac{1}{\epsilon}; m_{\alpha} \right)
\label{appeq:seven},
\end{align}
where $_1 F_1(a;b;x)$ and $_2 F_2(a_1,a_2; b_1,b_2;x)$ are Kummer's confluent hypergeometric function and the generalized hypergeometric function, respectively. When the mean photon number $ m_{\alpha}\gg1$, the contribution from $\delta P_{L;2}$ is negligible, and we focus on examining $\delta P_{L;1}$. The expression of $\delta P_{L;1}$ is greatly simplified for symmetric bare parameters $\Gamma=g$,
\begin{equation}
\delta P_{L;1} |_{\epsilon =1} = \frac{2}{m_{\alpha}} (1- e^{- m_{\alpha}})
\label{appeq:eight},
\end{equation}
showing a similar behavior to $\delta P_{L;1} |_{\epsilon =1}$ for a Fock state of photons at large photon numbers.

For a general $\epsilon$, we invoke the previously obtained asymptotic expansion of Kummer's function, \cite{kummer} 
\begin{align}
\delta P_{L;1} = \frac{2}{1+ \epsilon m_{\alpha}} \Big( & 1+\frac{m_{\alpha}}{(m_{\alpha}+\epsilon^{-1})^2} -\frac{m_{\alpha}}{(m_{\alpha}+\epsilon^{-1})^3} \nonumber \\
&+ \frac{3m_{\alpha}^2 + m_{\alpha}}{(m_{\alpha}+\epsilon^{-1})^4} + \cdots \Big)
\label{appeq:nine},
\end{align}
valid for $ m_{\alpha}\gg1$ and $\epsilon\ll1$. Keeping the first two terms in the bracket we obtain Eq.~(\ref{eq:seventeen}) in the main text. 

For a thermal state of photons, we start with Eq.~(\ref{eq:nineteen}) in the main text. The summations yield 
\begin{align}
\delta P_{L;1} & =  \frac{2}{m_{\beta}+1} \: _2 F_1  \left(1,\frac{1}{\epsilon}; 1+\frac{1}{\epsilon} ; \frac{m_{\beta}}{m_{\beta}+1} \right),  \nonumber \\
\delta P_{L;2} &=\frac{3/2}{m_{\beta}+1} \: _3 F_2  \left(1,\frac{1}{\epsilon}, \frac{1}{\epsilon}; 1+\frac{1}{\epsilon}, 1+\frac{1}{\epsilon} ; \frac{m_{\beta}}{m_{\beta}+1} \right)
\label{appeq:ten},
\end{align}
where $_2 F_1  (a_1,a_2; b;x)$ is the Gauss hypergeometric function and $_3 F_2  (a_1,a_2,a_3;b_1,b_2;x)$ is a generalized hypergeometric function. Eq.~(\ref{appeq:ten}) is easily verified using the series definitions of the hypergeometric functions. It is interesting to notice the likeness between these expressions and the response function of a particle crossing a quantum barrier while coupled to a thermal bath. \cite{wolynes1981} Whereas in the latter problem of dissipative tunneling the bath degrees of freedom were assumed weakly coupled to the particle for the convenience of analytical treatment, in our case the thermal photon mode dynamically couples to the electron and the process is dissipationless in an isolated system. 
At large photon numbers, $\delta P_{L;2}$ is negligible as in the case of a coherent state of photons. For symmetric bare parameters, we have
\begin{align}
\delta P_{L;1} |_{\epsilon =1} &= \frac{2}{m_{\beta}} \ln(m_{\beta}+1),  \nonumber \\
\delta P_{L;2} |_{\epsilon =1} &= \frac{3/2}{m_{\beta}} \: \mathrm{Li}_2 \left(\frac{m_{\beta}}{m_{\beta}+1} \right)
\label{appeq:eleven},
\end{align}
where $\mathrm{Li}_2(x)$ is the polylogarithm function of order 2. Eq.~(\ref{appeq:eleven}) manifests the logarithm factor in $\delta P_L$, a characteristic feature of a thermal state. 

To proceed, we obtain the asymptotic expansion
\begin{align}
\delta P_{L;1} = & \frac{2}{\epsilon(m_{\beta}+1)}   \sum_{n=0}^{\infty} \frac{\prod_{k=0}^n (1+k\epsilon)}{1+n \epsilon} \nonumber \\
& \times \; \frac{ \ln (m_{\beta}+1) - \psi(n+\epsilon^{-1}) +\psi(n+1)}{n! \; \epsilon^n(m_{\beta}+1)^n }   
 \label{appeq:twelve}
\end{align}
valid for $m_{\beta}\gg1$, where $\psi (x)$ is the digamma function. 
To see that Eq.~(\ref{appeq:twelve}) produces a sensible macroscopic limit, consider the lowest-order terms in the summation. Naively, $\ln (m_{\beta}+1) $ and $\psi (\epsilon^{-1})$ are singular when $m_{\beta}\rightarrow \infty$ and $\epsilon \rightarrow 0$. However, using the asymptotic expression of the digamma function with a large argument $x \gg 1$, \cite{expansion}
\begin{equation}
\psi(x+s) = \ln x+\sum_{n=1}^{\infty} \frac{(-)^{n+1}B_n(s)}{n} x^{-n} 
\label{appeq:thirteen},
\end{equation}
where $s$ can be a real number and $B_n(s)$ are Bernoulli polynomials, we find 
\begin{equation}
\ln (m_{\beta}+ 1)- \psi(\epsilon^{-1}) \simeq \ln \epsilon(m_{\beta}+ 1) + \frac{\epsilon}{2}
\label{appeq:fourteen}.
\end{equation}
One can check using Eq.~(\ref{appeq:thirteen}) that the higher-order terms are also regular in the macroscopic limit. 

When $\epsilon \gg 1$, one can use the series definition of the digamma function, \cite{expansion}
\begin{equation}
\psi(x) =-\gamma_0 - \frac{1}{x} +\sum_{n=1}^{\infty} \frac{x}{n(x+n)}
\label{appeq:fifteen},
\end{equation}
where $\gamma_0$ is Euler's constant, to show that Eq.~(\ref{appeq:twelve}) is still regular. However, in such a strong-coupling limit, \emph{e.g.}, when the electron-photon coupling strength $g$ is of a comparable order of magnitude with that of the photon energy $\omega$, it is known that the rotating-wave approximation in the Jaynes–Cummings model breaks down and the model becomes unphysical. In particular, the ground state of the system will no longer be one involving the vacuum state of photons. \cite{strong-coupling} 

Finally, when $\epsilon =1$, the summation reduces to a geometric series which recovers the expression in Eq.~(\ref{appeq:eleven}). We conclude that Eq.~(\ref{appeq:twelve}) is well-defined in a large domain of $\epsilon$. 

We emphasize here that $\delta P_{L;1}$ is a good approximation of $\delta P_{L}$ only at large photon numbers in a finite-size system, or at high light intensities in free space. Otherwise, one must resort to the exact results in Eqs.~(\ref{appeq:seven})(\ref{appeq:ten}) for an accurate estimation of $\delta P_{L}$. Keeping the lowest-order terms in the summation in Eq.~(\ref{appeq:twelve}), using Eq.~(\ref{appeq:fourteen}), and sending $\epsilon \rightarrow0$ for the macroscopic limit, we obtain Eq.~(\ref{eq:twenty}) in the main text.

%% For numbered reference style

\end{document}